\documentclass[fleqn,usenatbib]{mnras}

\usepackage{newtxtext,newtxmath}

\usepackage{sidecap}

\usepackage[T1]{fontenc}

\DeclareRobustCommand{\VAN}[3]{#2}
\let\VANthebibliography\thebibliography
\def\thebibliography{\DeclareRobustCommand{\VAN}[3]{##3}\VANthebibliography}

\usepackage[normalem]{ulem}

\usepackage{graphicx, caption, subcaption}	
\usepackage{amsmath}	

\usepackage[dvipsnames]{xcolor}

\title[How LSS shapes the MZR in time]{The Role of Large-Scale Environment in Shaping the Stellar Mass-Gas Metallicity Relation Across Time}

\author[A.R. Rowntree et al.]{Aaron R. Rowntree$^{1}$ \thanks{E-mail: a.rowntree-2018@hull.ac.uk (ARR)}, Fiorenzo Vincenzo$^{2}$, Ankit Singh$^{3}$, Changbom Park$^{4}$, Jaehyun Lee$^{5}$,\newauthor Christophe Pichon$^{4,6,7}$, Yohan Dubois$^{6}$, Gareth Few$^{1}$, Brad Gibson$^{8}$, Owain Snaith$^{9}$,\newauthor Yonghwi Kim$^{10}$
\\ ~ \\
$^{1}$E.~A. Milne Centre for Astrophysics, University of Hull, Hull, HU6 7RX, UK\\
$^{2}$Dipartimento di Fisica e Astronomia “Ettore Majorana”, Università degli Studi di Catania, Via S. Sofia 64, 95123 Catania, Italy\\
$^{3}$School of Physics and Astronomy, The University of Nottingham, University Park, Nottingham, NG7 2RD, UK\\
$^{4}$Korea Institute for Advanced Study (KIAS), 85 Hoegiro, Dongdaemun-gu, Seoul 02455, Republic of Korea\\
$^{5}$Korea Astronomy and Space Science Institute, 776, Daedeokdae-ro, Yuseong-gu, Daejeon 34055, Republic of Korea\\
$^{6}$Institut d’Astrophysique de Paris, UMR 7095, CNRS, Sorbonne Université, 98 bis boulevard Arago, 75014 Paris, France\\
$^{7}$Kyung Hee University, Dept. of Astronomy and Space Science, Yongin-shi, Gyeonggi-do 17104, Republic of Korea\\
$^{8}$Woodmansey Primary School, Hull Road, Woodmansey HU17 0TH, UK\\
$^{9}$University of Exeter, School of Physics and Astronomy, Stocker Road, Exeter, EX4 4QL, UK \\
$^{10}$Korea Institute of Science and Technology Information (KISTI), 245 Daehak-ro, Yuseong-gu, Daejeon 34141, Republic of Korea
}

\date{Accepted 2026 Jan 23; Received 202- xxx; in original form 202- xxx}
\pubyear{202-}

\begin{document}

\maketitle

\begin{abstract}

    We study the stellar mass-gas metallicity relation (MZR) which shows a significant scatter for a fixed stellar mass. By defining global environments, nodes, filaments, and voids within the Horizon Run 5 cosmological hydrodynamical simulation, we explore when and where the enrichment of galaxies occurs, analysing key evolution parameters such as star-formation rate and changes in gas-fraction and gas-metallicity per unit time. At high redshift ($z>4.5$), there are minimal deviations from the MZR due to environment, however, larger deviations emerge as redshift decreases. Low stellar mass galaxies in nodes, $M_{\star} < 10^{9.8}\,\text{M}_{\odot}$, start showing deviations at $z = 3.5$, whilst other environments do not. For, $z < 2$, filaments and voids begin to show deviations above and below the MZR, respectively. By $z = 0.625$, the last epoch of HR5, deviations exist for all stellar masses and environments, with a maximum value of 0.13 dex at $M_{\star} \approx 10^{9.35}\,\text{M}_{\odot}$, between the median gas metallicities of node and void galaxies. To explain this environmental variance we discuss gas accretion, AGN, ram-pressure-stripping and strangulation as regulators of $Z_{g}$. Concurrently, at high metallicities, for $z < 2$, while massive galaxies in nodes show increasing $Z_{g}$ and decreasing [O/Fe], void galaxies show a turnover where $Z_{g}$ falls with decreasing [O/Fe]. This directly points to the importance of cold-gas accretion in retaining lower $Z_{g}$ in massive void galaxies for $z < 2$, whilst its absence in nodes allowed $Z_{g}$ to access higher values.
    
\end{abstract}

\begin{keywords}
cosmology: large-scale structure -- galaxies: formation -- galaxies: evolution -- galaxies: kinematics and dynamics --  galaxies: high-redshift -- methods: numerical 
\end{keywords}

\section{Introduction}
\label{sec:intro}

The structure observed in our Universe is a truly multi-scale and multidimensional feature. On the largest scales, the evolution of structure since the cosmic dawn led to the formation of the so-called Cosmic Web \citep{Bond1996}. Underpinned by dark matter, the Cosmic Web hosts a network of interconnecting filaments that join the massive, gravitationally bound galaxy clusters located at the nodes of the web \citep{Gregory1978}. Between these bright regions is an expanse of almost empty voids \citep{Kirshner1987}. The density contrast observed in the local universe is expected to be initially seeded by pre-inflationary quantum fluctuations, which eventually lead to the gravitational collapse of matter in these regions \citep{Zeldovich1970}. Early in its history, the Universe was dominated by a larger number of spurious filaments and sheets with low density-contrasts. Subsequent evolution involving the continuous accretion of matter and the merging of these structures led to the high-density contrasts that emerge at low redshift \citep{Cautun2014}. It has been shown that there is an intrinsic relationship between these structures, with matter flowing from voids into and along large-scale filaments, eventually ending its journey at the nodes, facilitating the evolution of galaxies along this path \citep{1997Dressler, 2004Kauffmann, 2012Peng, 2016Boselli}. Observational studies have shown that galaxies and certain galaxy properties are related to the environment that they exist within, such as star formation rate (SFR) \citep{Porter2008, Peng2010, Haines2011, 2016Alpaslan, 2016Mart,  Mahajan2018, 2020Singh, Gallazzi2021}, gas-fraction \citep{Hasan2023}, colour \citep{Hoosain2024} and metallicity \citep{Shields1991, Henry1992, Cooper2008, Donnan2022}. Similar conclusions were obtained using large-scale cosmological simulations \citep{Metuki2015,Gupta2018, 2020Singh}. Various mechanisms such as galaxy-mergers \citep{2006Kewley,2015Sobral,2015Stroe,LHullier2012}, ram pressure stripping \citep{1972Gunn,2019Singh, Jhee2022, 2024Singh} and harassment \citep{1996Moore} have been proposed to explain such correlation between the galaxy properties and the environment a galaxy exists within. \par
Metallicity provides significant insight into how a galaxy evolves, providing a historical record of its chemical enrichment and the dynamic processes that have led to its current state \citep{Mernier2020}. Extensive studies have shown metallicity to be linked to inflows, stellar winds, active galactic nuclei (AGNs), supernova (SN) explosions, stellar mass, star-formation rate (SFR), and gas fraction. Many of these processes and properties have also been related back to the environment. The study of metallicity comes hand in hand with the mass-metallicity relation (MZR) \citep{Tremonti2004}, linking stellar mass and metallicity to show a strong positive correlation between the two properties. However, at a fixed stellar mass, the MZR shows a significant scatter in metallicity, suggesting that other galaxy properties or processes also drive the chemical enrichment of galaxies. 
Over the years, the MZR and its scatter have been the focus of a large number of studies looking at samples of galaxies in the local Universe \citep{1979Lequeux, Tremonti2004} and at higher redshifts \citep{2005Savaglio, 2006Maier, 2008Maiolino, 2012Foster, 2013Zahid, 2013Moller}. The MZR has also been shown to have its own dependency on the environment \citep{Thomas2005, Yates2012, 2013Pilyugin, 2013Sanchez, 2014Peng, Donnan2022, Rowntree2024}. 
Early key studies found higher chemical enrichment for Virgo group galaxies when compared to the field \citep{Shields1991, Henry1992, Skillman1996}. Using Sloan Digital Sky Survey (SDSS) data, \citet{2009Ellison} reported metal enrichment above the field for cluster galaxies, independent of galaxy size and cluster properties. \citet{2017Wu} found a relationship between the MZR and local density, presenting a slight dependence of the MZR on the environment in this context. More recently, \citep{Cedres2024}, using the GLACE survey, also showed that the MZR for galaxies within clusters is enhanced above the average, beyond this, larger mass clusters show even more enrichment.
Understanding the vertical scatter in the MZR is also of great importance when considering the general evolution of galaxies, as it demonstrates the importance of underlying processes beyond the MZR. Within the last two decades, significant developments have been made in this area. In observations at $z = 0.7$, \citet{Choi2014} measured the stellar MZR, finding differing average stellar metallicities for galaxies when compared to the median measured MZR for the same redshift, which further indicates a large scatter within the population. Other studies using cosmological-hydrodynamical simulations dug deeper into this topic, with \citet{Torrey2019} finding the scatter in the MZR to strongly correlate with gas mass and SFR and \citet{vanLoon2021} clarifying the dependencies on gas-fraction, inflow rate and outflow rate. In recent works like  \citet{Donnan2022}, \citet{Dominguez2023} and \citet{Rowntree2024}, the link back to the large-scale structure in the Universe is made, providing a clear connection between the MZR and cosmic environments. In all studies, an agreement is found that nodes or galaxies in clusters show significant enrichment compared to the total MZR, whilst galaxies in the field, or void, show under enrichment. 
On top of an environmental dependency, it is observed that over redshift, the MZR has its own time-evolution, showing an increasing metallicity at all stellar masses with the passage of time \citep{2008Maiolino, Langeroodi2023}.
As it is clear that these aforementioned global environments and their density-contrasts are dynamic in redshift, one can intuitively expect that the relationship between the MZR, its scatter, and the environment also depends on redshift. This study aims to build on this question by investigating how the evolution of the MZR is affected by three unique cosmic environments: nodes, filaments, and voids from redshifts $z = 0.5$ to $4.5$. 
There is a reasonable amount of recent observational work that studies the impact of local environment on gas-phase metallicity. \citealt{Chartab2021} uses the spectra from the near-IR, MOSFIRE deep-evolution-field survey to calculate the gas-phase oxygen abundances of galaxies between $z = 1.4$ and $2.6$. The survey also provides photometric redshifts that the study uses to calculate local environment measurements. As expected, at the low end of their redshift range, gas-phase metallicity is enhanced for galaxies in local over-densities, however interestingly, for higher redshifts, this trend is reported to be reversed. \citealt{Calabro2022}, with a similar methodology, also finds that between $z = 2$ and $4$, using a combined VANDELS and MOSFIRE dataset, gas-phase metallicity shows a noticeable reduction for galaxies in the largest local over-densities. What is common in current observational studies is the basis of local environment, typically based on the number of galaxies in a specific volume centred on each galaxy, which differs greatly from the identification and definition of specific global environments, i.e. nodes, filaments and voids, or a statistical measure, like nearest neighbour calculations, that are based in different quanta and represent different physical contexts. Unfortunately, in observations the definition of global environments must be based on the galaxies as matter fields, DM or otherwise, are not available. As such, a survey must have sufficient volume, for structure estimation, and spectroscopic accuracy, for the measurement of gas-phase metallicity, to truly probe the impact of these global environments on chemical enrichment. This process has been emulated in similar work based on simulations, using the galaxies to trace the large-scale structure, basing the studies in a space more comparable to observations. This is what we also aim to achieve with this work. Until the eventual arrival or full data release of the new generation of galaxy surveys (for example, the Dark Energy Spectroscopic Instrument, \citealt{DESI2023}), acquiring accurate spectroscopy whilst concurrently measuring the large-scale structure based on the galaxies is not particularly achievable. This paper aims to provide useful insight into when and where galaxies become enriched in the context of the large-scale structure, building on why this may be, thereby providing a meaningful groundwork to inspire further data analysis using upcoming observational data. In this work, we use the cosmological-hydrodynamical simulation, Horizon Run 5 (HR5) \citep{Lee2021}, which we present an overview of in the following section. HR5 allows us to tackle both Mpc and kpc scales concurrently, allowing the link between the galaxy properties building the MZR and the over-arching large-scale structure to be placed in the same frame.

This study is structured as follows. Section \ref{sec:Method} outlines the HR5 simulation, our data preparation and our methodology used to compute the individual cosmic environments. The results of this method are presented in section \ref{sec:Results} and are discussed in section \ref{sec:Discussion} , with our main conclusions left in section \ref{sec:Conclusions}.

\section{Method}
\label{sec:Method}

In this work, we use the Horizon Run 5 (HR5) cosmological hydrodynamical simulation \citep{Lee2021} that models physical processes from kpc to Mpc scales within a cosmological volume using an {\tt MPI-OpenMP} version of the adaptive mesh refinement (AMR) code {\tt RAMSES} \citep{Teyssier2002}. The simulation runs from $z = 200$ down to $z = 0.625$ where the simulation halts. HR5 full box spans a volume of $1.049 \times 1.049 \times 1.049$ $\text{cGpc}^{3}$. Within this, a smaller cuboid region of $1049 \times 119 \times 127$ $\text{cMpc}^{3}$ is initially set as the highest resolution volume, in which grids can be refined down to $\Delta x=1\,$kpc, depending on local density. Over the full timescale of the simulation, a range of sub-grid physics is employed each of which contributes to the overall evolution of the galaxies within: supernovae (SN) (\citealt{Dubois2008}), AGN feedback (\citealt{Dubois2012}), star-formation activity (\citealt{Rasera2006}), ultraviolet background heating (\citealt{Haardt1996}) and metallicity-dependent radiative cooling (\citealt{Dalgarno1972}). The chemical evolution of galaxies within HR5 is governed by the outputs of RAMSES-CH \citep{Few2012}, which employs a continuous feedback model. It allows energy and chemical feedback to be followed through AMR, providing constraints on abundance ratios and supernova rates. Note that the specific redshifts and mass scales mentioned in this study, with reference to the HR5 simulation, should only be interpreted qualitatively as they are sensitive to the simulation parameters. As such, results are only based on relative redshift evolution and mass dependence. HR5 follows the standard $\Lambda$ cold dark-matter ($\Lambda$CDM) Universe with the following cosmological parameters: $\Omega_{0}=0.3$, $\Omega_{\Lambda}=0.7$, $\Omega_{\rm b} =0.047$, $\sigma_{8} = 0.816$, and $H_{0}= 100\times h_{0}=68.4\,\text{km}\,\text{s}^{-1}\,\text{Mpc}^{-1}$, which are compatible with the Planck data \citep{Plank2016}.

\subsection{Galaxy distribution and selection}

HR5 provides galaxy and halo catalogues at 128 snapshots between $z = 15.555$ and $z = 0.625$. Galaxy catalogues contain extensive properties that fully describe each galaxy in every snapshot. They also contain numerous flags that allow easy tracking of individual galaxies between snapshots to begin to study their evolution in time. For a short summary of how these catalogues are populated, we refer the reader to \citealt{Rowntree2024}, and for a more detailed description, see \citealt{Lee2021, Kim2023}.

This work aims to study the evolution of galaxies over redshift in different cosmic environments, and as such, we firstly select 9 main snapshots between $z = 4.5$ and $z = 0.5$, which are outlined in the Table. \ref{table1}. The volumes provided for each snapshot are the zoomed-in regions that are not contaminated by low-level dark matter particles. The size of this region decreases with a falling redshift due to the evolution of the matter density fields.

\begin{table}
    \centering
    \caption{The snapshot selection taken from the HR5 simulation, showing each snapshots redshift, its reduced, uncontaminated co-moving volume and the number of galaxies within it.}
    \begin{tabular}{ |c|c|c| }
     z & Volume [$\mathrm{10^{7} cMpc^{3}}$] & \# of Galaxies \\ 
     \hline
     $\approx 0.5$ & 1.282 & 498508 \\  
     \hline
     $\approx 1$ & 1.290 & 534913 \\
     \hline
     $\approx 1.5$ & 1.301 & 545764 \\
     \hline
     $\approx 2$ & 1.309 & 539537 \\
     \hline
     $\approx 2.5$ & 1.316 & 497989 \\
     \hline
     $\approx 3$ & 1.322 & 431128 \\
     \hline
     $\approx 3.5$ & 1.327 & 335165 \\
     \hline
     $\approx 4$ & 1.331 & 261462 \\
     \hline
     $\approx 4.5$ & 1.335 & 198324 \\
     \hline
    \end{tabular}
    
    \label{table1}
\end{table}

    These raw galaxy catalogues require two main reductions or cuts to arrive at a suitable dataset for our analysis. We follow a process for each catalogue identical to \citealt{Rowntree2024}. The first is a removal of galaxies that lie near low-resolution dark matter (DM) regions. Lastly, we remove all galaxies with stellar mass, $M_{\star}< 2 \times 10^{9}\,\text{M}_{\odot}$, to ensure that each galaxy is at least resolved with $\sim 10^3$ stellar particles. For a more detailed description of these data reductions, we refer to \citet{Rowntree2024}. This process is carried out exactly for each of the nine snapshots considered in this study.

    \subsection{Quantification of cosmic structure}

    To define the cosmic environments of galaxies, we provide a quantitative description of the large-scale structures of the universe. Our work focuses mainly on three cosmic environments: nodes, filaments, and voids. There are many ways to provide a robust identification of these environments; one common example is filament-finding algorithms. These algorithms, namely {\tt DisPerSE} \citep{Sousbie2011}, {\tt Nexus} \citep{Cautun2013} and {\tt T-ReX} \citep{Bonnaire2020, Bonnaire2022}, trace the large-scale structures in the input distribution, galaxies or DM, and output a skeleton-like distribution of filaments and nodes that have a quantitative position in the volume. Each of these separate algorithms makes use of a different methodology, for example, using the topology of the matter density field \citep{Sousbie2011}, Hessian matrices \citep{Cautun2013}, regularised minimum spanning trees \citep{Bonnaire2020}, and recently machine learning methods \citep{Inoue2022, Awad2023}. For a review and comparison of these individual algorithms, see \citet{Libeskind2018}. In this work, we use {\tt T-ReX} \citep{Bonnaire2020, Bonnaire2022}. This algorithm is purpose-built to detect filamentary structure in observational surveys within which the information describing the structure is contained in the galaxy-point distribution; as such, we are utilising the galaxies as the tracer of the underlying structure \citep{Galarraga2020, Galarraga2023, Gouin2021, Donnan2022, Bulichi2023}. Using galaxies as the tracer of the DM density field is a biased methodology which we outline in \citealt{Rowntree2024}. Although biased, we place importance in being comparable to future observational work that typically will be working from a similar methodology.
    The structure estimated by {\tt T-ReX} is produced by a combination of Gaussian mixture models (GMM) describing the distribution of the tracers and graph theory, more specifically, a regularised minimum spanning tree, to produce a smooth connection between the tracers \citep{Bonnaire2020}. The final output from {\tt T-ReX} is a list of edges in space that, when combined, create a `skeleton' that estimates the large-scale structure present in the distribution of tracer particles.

    To use {\tt T-ReX} we must first prepare our input data to ensure that we provide the algorithm with the correct information to create structure estimates from. It is common to only select the high stellar mass galaxies as it has been shown in the literature that these galaxies are more likely to lie on the structure itself \citep{Sarron2019, Galarraga2020}. However, due to the effects shown in appendix A of our previous paper \citep{Rowntree2024}, we again opted out of using this methodology. Due to the consideration of redshift evolution in this work, we must further develop the method used in \cite{Rowntree2024}. As the Universe evolves, the galaxy distribution evolves with it, such that there is a significantly different number of galaxies within different volumes at different redshifts. As such, without careful consideration, the created skeletons will not be statistically comparable and will not be probing the same scale structures. For example, a skeleton computed at a redshift of 10, with a low galaxy density, may lead to longer and more spurious structures, whilst a low redshift, high galaxy density sample may lead to smaller scale filamentary structure being identified. Therefore, the types of filament detected by these two skeletons are incomparable, exist on different spatial scales, and are physically different categories of structure. The goal is to ensure that the structure that {\tt T-ReX} returns from each snapshot is probing similar length scales, making the structure observed at one time comparable to that observed at another time. It was recently achieved in \citet{Galarraga2024} by selecting the same number of galaxies at each snapshot in MillenniumTNG, leading to a consistent density when computing the skeletons. The main difference in our work is that HR5 is a zoom-in simulation. The surface of the HR5's zoom-in volume can be contaminated by low-resolution massive particles at later epochs that originated outside the initial zoom-in region but have permeated into it with time. To account for this, the `pure' co-moving volume of this zoom-in region is measured at each snapshot to ensure that the region we study remains unaffected by these permeating low-resolution particles. Our new methodology considers this and has elements similar to the previous one; however, the following section will outline the full method to ensure clarity. \par 
    
    To select the galaxies for skeleton computation, we start from the full catalogue with the impure and low-resolution galaxies removed. Rather than taking a random selection of these galaxies as done in \cite{Rowntree2024}, in this work, we calculate the local number density at the location of each galaxy, $\rho_{g}$, as $\rho_{g} = 1 / d_{10}$, where $d_{10}$ is the distance to the tenth closest galaxy with $M_{\star}> 2 \times 10^{9}\,\text{M}_{\odot}$. To account for the structure's redshift evolution and ensure the comparability of our skeletons between redshifts, we also fix the galaxy density in each volume given to {\tt T-ReX} for each skeleton computation. First, we select a constant galaxy density $\rho_{i}$ to set at each snapshot. It is obtained by calibrating the galaxy density underneath the skeletons at different snapshots using the $\rho_{g}$ distribution at $z = 0.5$; at this redshift with pure volume, $V_{0.5}$, and galaxy count, $N_{0.5}$, we select the top 50\% of galaxies in $\rho_{g}$, which leads to a galaxy number density $\rho_{i} = 0.02519 \, \mathrm{cMpc^{-3}}$. This $\rho_{i}$ value, equivalent to a mean separation, $\overline{d} = 4.9 h^{-1}\mathrm{cMpc}$, is consistent with the SDSS number density of galaxies with r-band absolute magnitude brighter than roughly $-19.8+5 \mathrm{\log(h)}$, where $h$ is the reduced Hubble constant, $h = H_{0}/(100 \mathrm{[km s^{-1} Mpc^{-1}]})$ \citep{Croton2013}. This is 0.5 magnitudes fainter than the SDSS cutoff absolute magnitude $M_{\star}$ in the r-band luminosity function at redshifts $0.025 < z < 0.010713$ \citep{Choi2007}. We then calculate the percentile of $\rho_{g}$ required in each new snapshot with galaxy count $N_{i}$ in volume $V_{i}$ that reproduces the same $\rho_{i} = 0.02519 \, \mathrm{cMpc^{-3}}$. By repeating this for all snapshots, we have datasets with the same galaxy density. As such, the skeletons computed from these datasets are based on the same spatial scales and are, therefore, more comparable. \par 
    
    To further ensure the comparability and, more so, the robustness of our skeletons, we turn to two properties: the filament length, $l_{\text{fil}}$, distribution, and the galaxy-to-skeleton distance, $d_{\text{skel}}$, distribution. The filament length distribution explicitly shows whether the skeletons and the filaments associated with them are within the same length scales at different redshifts. The $d_{\text{skel}}$ distribution is another check to ensure there are no significant spurious spatial differences in how galaxies are distributed around the skeletons. Both distributions are expected to be uni-modal, as shown in \citet{Rowntree2024}. To calculate $d_{\text{skel}}$, we make use of the {\tt radial\_distance\_skeleton} function, from within the {\tt T-ReX} module, that calculates the perpendicular distance from each galaxy to its nearest edge on the skeleton. We execute the function between every galaxy selected from each snapshot and the corresponding skeleton at that redshift, leading to a $d_{\text{skel}}$ value that represents the perpendicular distance from galaxy to skeleton, for every galaxy at each redshift. We compare these distributions with our previous work and similar work in the field.

    We manually tune {\tt T-ReX}'s three main parameters with the density $\rho_{i}$ to find the parameter set that produces the most robust skeleton by our defined assessments. These parameters are: \textit{(i)} $\Lambda$, the length and smoothness of the skeleton; \textit{(ii)} $l$, the de-noising parameter, and \textit{(iii)} $\sigma$, which alters the variance of the GMMs that are in use, \citep{Bonnaire2020}. In slightly more detail, $\Lambda$ acts as a soft-constraint on the overall length of the skeleton, where a larger value leads to a shorter and smoother skeleton. It equally helps with robustness, maintaining the tree-like structure in noisy or under-dense regions. $l$ more accurately refers to a level in the minimum spanning tree that beyond which, extremities of the skeleton are removed. This retains the core shape of the skeleton, whilst removing spurious outer edges. Finally, $\sigma$ controls the variance and extent of the Gaussian clusters. This is another tunable parameter that impacts the coarseness or smoothness of the skeleton. A small value produces a more coarse skeleton, whilst a large value helps with higher smoothing. These parameters must be tuned for a specific galaxy density within a volume to ensure a statistically reasonable estimate of the structure is produced. As we manually fixed the number density of galaxies at $\rho_{i}$ within each volume at each redshift, we use the same parameter set for every snapshot. To begin the tuning process, we test the parameter set used in \citep{Rowntree2024}. As the context is different, we did not expect this to be correct, and is merely used as a starting point. The key features we aim for in the produced PDFs of $l_{\text{fil}}$ are that the distributions are unimodal, and the PDF at each snapshot lies within the same spacial scales as one another, being comparable to that seen for the $l_{\text{fil}}$ function in \citep{Galarraga2024}. This initial test produced PDFs in $l_{\text{fil}}$ that lay in different scales for each snapshot and were bimodal, and PDFs in $d_{\text{skel}}$ that were also bimodal. By now altering a single parameter at a time, we could determine the impact of each parameter in moving the distributions closer to what is used in previous studies. By populating a 3d-grid of parameters centred on our initial choices, varying by ±$100\%$, maintaining correct signage, we found no success in producing a reasonable set of skeletons. However, using larger values of $\Lambda$ and $l$, concurrently with smaller values of $\sigma$ produced the most promising PDFs. By then moving the range of our parameter space in each axis to, $20 \leq \lambda \leq 40$, $0.3 \leq \sigma \leq 0.4$ and $20 \leq l \leq 30$, in steps of 10, we found agreeable outputs with a handful of the parameter sets. With a number of viable choices, each producing unimodal PDFs in both $l_{\text{fil}}$ and $d_{\text{skel}}$, within a reasonable filament length scale, we chose the set that produced PDFs that were closest in minimum, maximum and median filament length to the existing studies. The parameter set that we arrived at is $\Lambda = 30$, $\sigma = 0.38$, $l = 25$. Using this methodology and parameter selection, fixing the number density of galaxies within each volume at $\rho_{i}$ by choosing the corresponding upper percentile in the local density distribution of the galaxies, we produce nine skeletons ranging from $\mathrm{z} = 0.5$ to $4.5$.
    
\begin{figure*}
    \centering
    \includegraphics[scale=0.65]{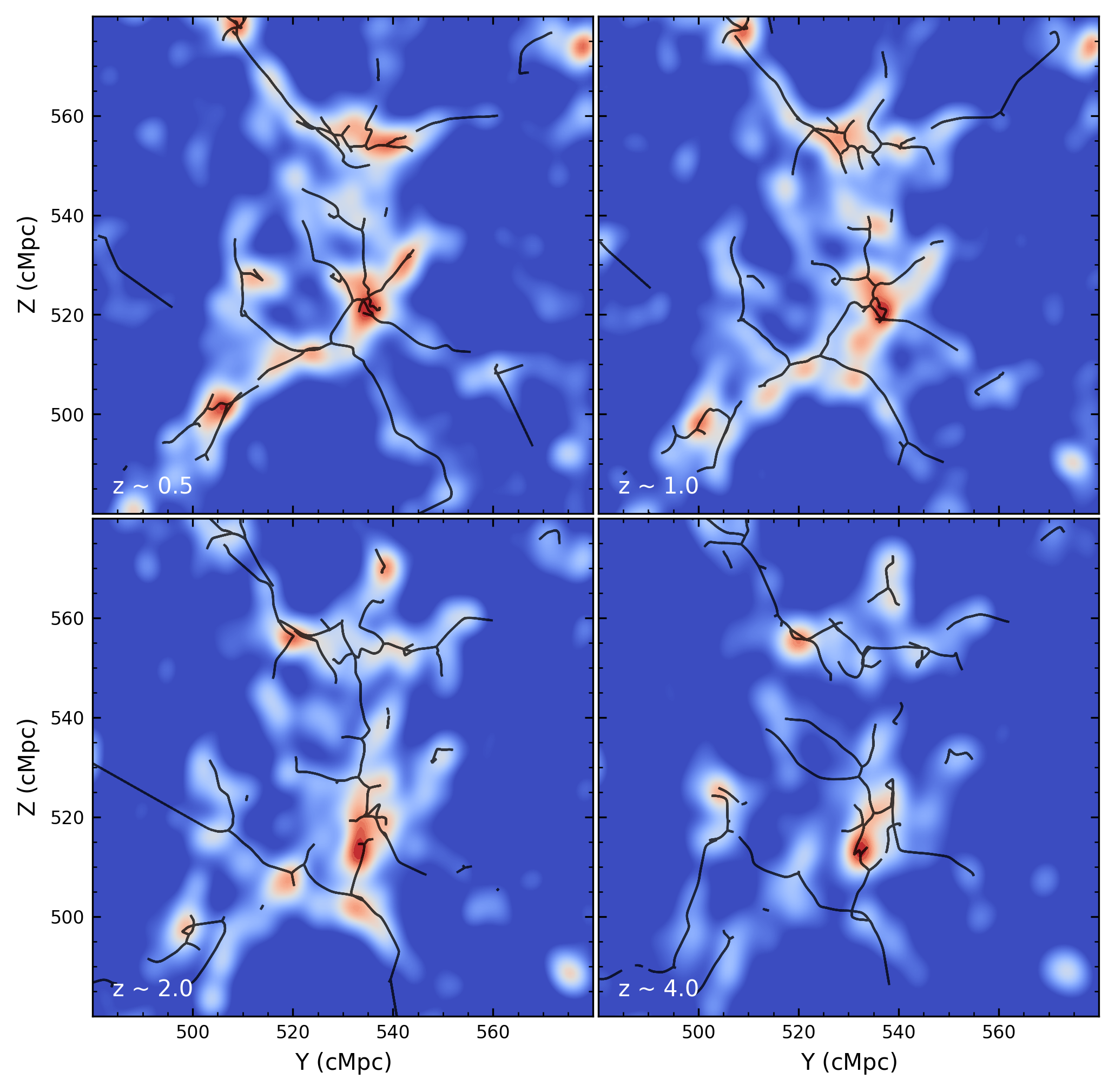}
    \caption{A 20 cMpc thick slice of the galaxy distribution in HR5 with the skeletons computed using {\tt T-ReX} for 4 different redshift snapshots at $z = 0.5$ (top left), $1$ (top right), $2$ (bottom left) and $4$ (bottom right), overlayed (black lines). The coloured contours represent a 2d-kernel density estimate of the galaxy distribution, with the more dense regions in red and less dense regions in blue used to qualitatively assess the skeleton.}
    \label{fig:Skeleton_Comparison}
\end{figure*}

\begin{figure*}
    \centering
    \includegraphics[scale=0.65]{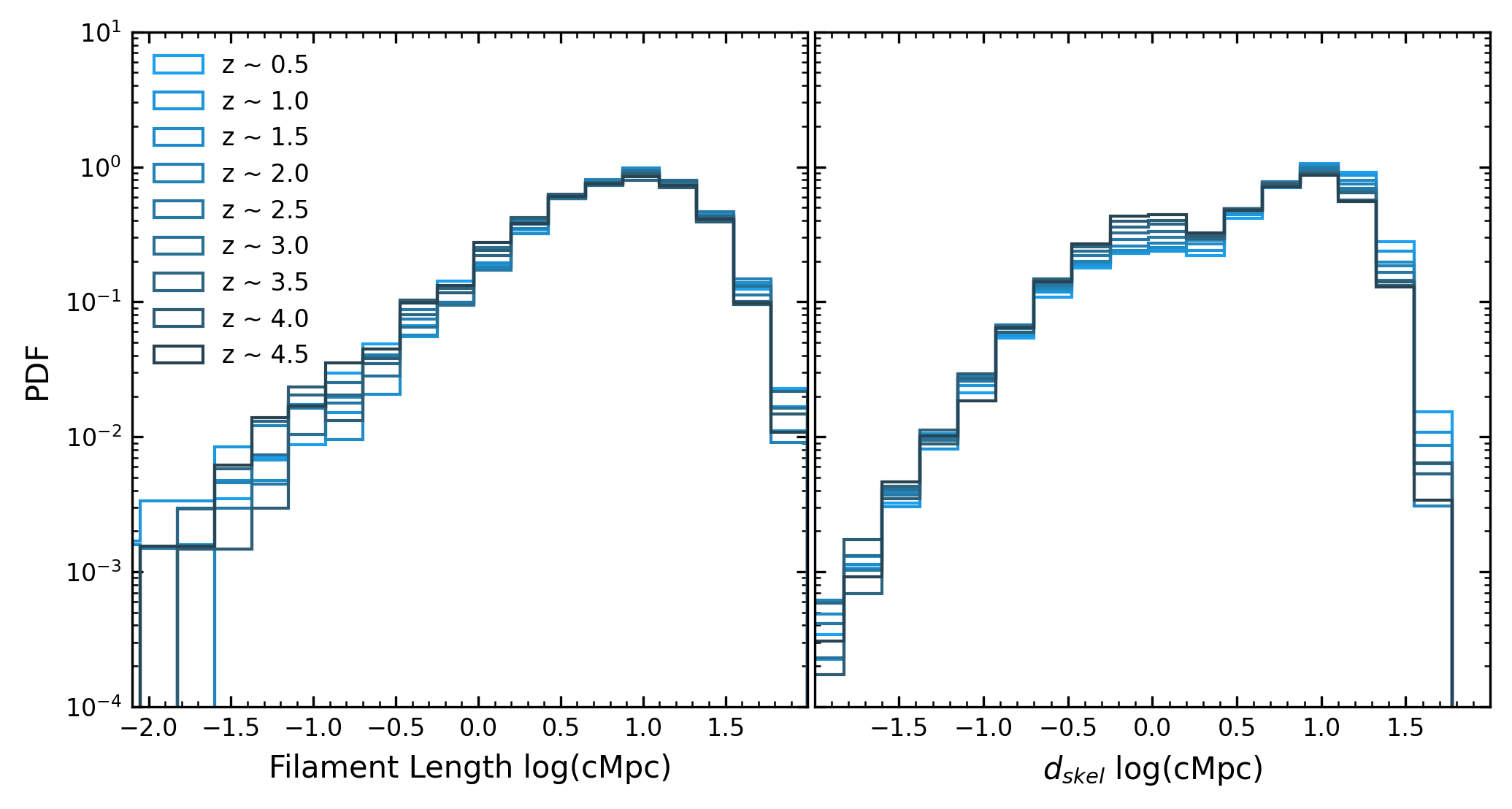}
    \caption{Probability density functions in filament length (left) and $d_{skel}$ (right) computed using skeletons at varying redshifts in HR5. Each 1-d histogram contains 20 equal-sized bins between $-2.5$ and $2$ log(cMpc). The lighter blues represent a lower redshift, while the darker blues represent a higher redshift, as seen in the legend.}
    \label{fig:Fil_Length_Dskel_PDFs}
\end{figure*}

    Fig. \ref{fig:Skeleton_Comparison} shows four of the nine skeletons in a $20 \, \mathrm{ cMpc}$ thick slice in HR5. The blue-to-red contour represents a kernel-density estimate of the selected galaxies, with the red regions being the most dense and the dark blue representing the least dense. The black overlaid lines demonstrate the skeleton computed by {\tt T-ReX}. On a purely visual, qualitative basis, {\tt T-ReX} is doing a reasonable job of tracing the higher-density regions.
    Fig. \ref{fig:Fil_Length_Dskel_PDFs} presents a quantitative measure of the robustness of the 9 skeletons. In the first panel, we see the probability density functions (PDFs) of the filament lengths in a histogram for the 9 snapshots overlayed with one another. All PDFs in filament length cover the same spatial range, suggesting the structures detected in each snapshot are of the same nature and scale. A similar conclusion can be drawn from the right panel, in which the $d_{\text{skel}}$ distribution is very similar for all redshifts, again covering the same spatial range. Combined, these two plots indicate that our computed skeletons are probing the same category of structure across cosmic time, allowing us to make meaningful comparisons on how galaxies inside or outside these structures evolve with redshift.

\subsection{Environmental definitions} 

We use the computed skeletons to define the specific cosmic environments that we consider in this work. In decreasing order of density contrast, we focus on three main environments: Nodes, Filaments, and Voids. This section will provide an overview of the three environments and the methods used to select the galaxies within them across the different snapshots.

\subsubsection{Nodes}

To identify Nodes in a particular snapshot, we use the largest galaxy clusters in the simulation as a proxy. Typically, these galaxy clusters will lie at the intersection between multiple filaments and, as such, are commonly the nodes of the skeleton. The halo catalogues contain information on total cluster mass, the centre-of-mass coordinates, $R_{200}$ value, flags for satellite and central galaxies, and a flag labelling whether the cluster is in the aforementioned well-resolved zoom-in region. We remove clusters lying too close to the low-resolution DM regions using this 'pure' flag. Following this we apply a cut in total mass, $M_{\text{tot}} \geq 10^{13}\,\text{M}_{\sun}$, at $z = 0.5$, selecting the largest galaxy clusters in the final snapshot. As the universe evolves, the large galaxy clusters accumulate more matter, shifting the median cluster mass towards higher values, showing the expected mass evolution of halos. As such, the cluster mass distribution is redshift dependent and therefore using a static $M_{\rm tot}$ cut at $10^{13}\,\mathrm{M_{\sun}}$ retrieves a different population of clusters at a different time in the universe. To account for the mass evolution of these structures, we use the mass cut of $M_\mathrm{tot} \geq 10^{13}\,\mathrm{M_{\sun}}$ in the last snapshot at $z = 0.5$ to calculate the percentile of FoF halos that this mass cut selects. This is equivalent to taking the top $0.03\%$ of halos in $M_{\rm tot}$. In the other eight snapshots, we proceed to take the top $0.03\%$ of halos in $M_{\rm tot}$ to ensure that we are selecting only the most massive structures at the specific redshift as our node proxies. See table \ref{table2} for the specific cluster mass cut-offs. A final cut then ensures that our sample of galaxy clusters is only composed of virialized clusters. By calculating the distance, $d_{\rm BCG}$, between each cluster's center of mass and the position of the brightest central galaxy, normalizing it to each cluster $R_{200}$ value, we can define a normalized offset, 
$\Delta_{r} = d_{\rm BCG}/R_{200}$. Note that the $R_{200}$ value used in this case represents the radius at which the mean matter density is 200 times the critical density of the Universe. A large offset suggests an unrelaxed system; for more detail on the calculation and significance of this property, refer to \citet{Rowntree2024}. We limit our selection of galaxy clusters in every snapshot to clusters with $\Delta_{r} < 0.05$. With this selection, we define any galaxy lying within $d_{\rm cluster } \leq 2 \times R_{200}$ of one of these clusters as a `node galaxy', with $d_{\rm cluster}$ representing its distance to the center of mass of the cluster it is within.

\begin{table}
    \centering
    \caption{The cluster total-mass cut-off used to define the Node population in each snapshot selected from HR5.}
    \begin{tabular}{ |c|c| }
     z & Cluster Mass Cut-off $\mathrm{[M_{\odot}}]$  \\ 
     \hline
     0.5 & $10^{13.00}$ \\  
     \hline
     1.0 & $10^{12.89}$ \\
     \hline
     1.5 & $10^{12.73}$ \\
     \hline
     2.0 & $10^{12.61}$ \\
     \hline
     2.5 & $10^{12.45}$ \\
     \hline
     3.0 & $10^{12.31}$ \\
     \hline
     3.5 & $10^{12.16}$ \\
     \hline
     4.0 & $10^{12.08}$ \\
     \hline
     4.5 & $10^{12.01}$ \\
     \hline
    \end{tabular}
    
    \label{table2}
\end{table}

\subsubsection{Filaments}

The membership of filament population is defined by the $d_{\rm skel}$ value of each galaxy. We calculate $d_{\rm skel}$ for all galaxies outside of $2 \times R_{200}$ of any galaxy cluster, i.e. galaxies that do not belong to nodes. A galaxy is defined to be in filament if it has $d_{\rm skel} \leq 1 \,\mathrm{cMpc}$. This value of $1 \,\mathrm{cMpc}$ is inspired by the typical filament radii of $1-3 \,\mathrm{Mpc}$ that are reported in literature \citep{Gheller2015,Galarraga2022, Wang2024}. To probe the cores of these structures, we set our radial cut at the low end of this range \citep{Galarraga2023}.
Due to the pre-existing consideration of the effect of redshift on the skeleton computation, we approach the definition of filament galaxies in the same way in every snapshot with these galaxies defined by $d_{\rm skel} \leq 1\,\mathrm{cMpc}$ relative to the skeleton computed at each redshift.

\subsubsection{Voids}
    
The void population follows simply from the filaments and again uses $d_{\rm skel}$ explicitly. A galaxy with $d_{\rm skel} \geq 8 \, \mathrm{cMpc}$ is considered a void galaxy. The cut in distance is the same as seen in \citep{Rowntree2024}, which was inspired by the distribution of galaxies in $d_{skel}$. For a detailed cover of this, refer to \citealt{Rowntree2024}. As mentioned for the filaments, since these environments are based on the skeletons computed for each redshift, the void definition stays constant across the snapshots.

\subsection{MZR residual calculation}

\begin{figure*}
    \centering   
    \hspace*{0.5mm}
    \includegraphics[scale=0.627]{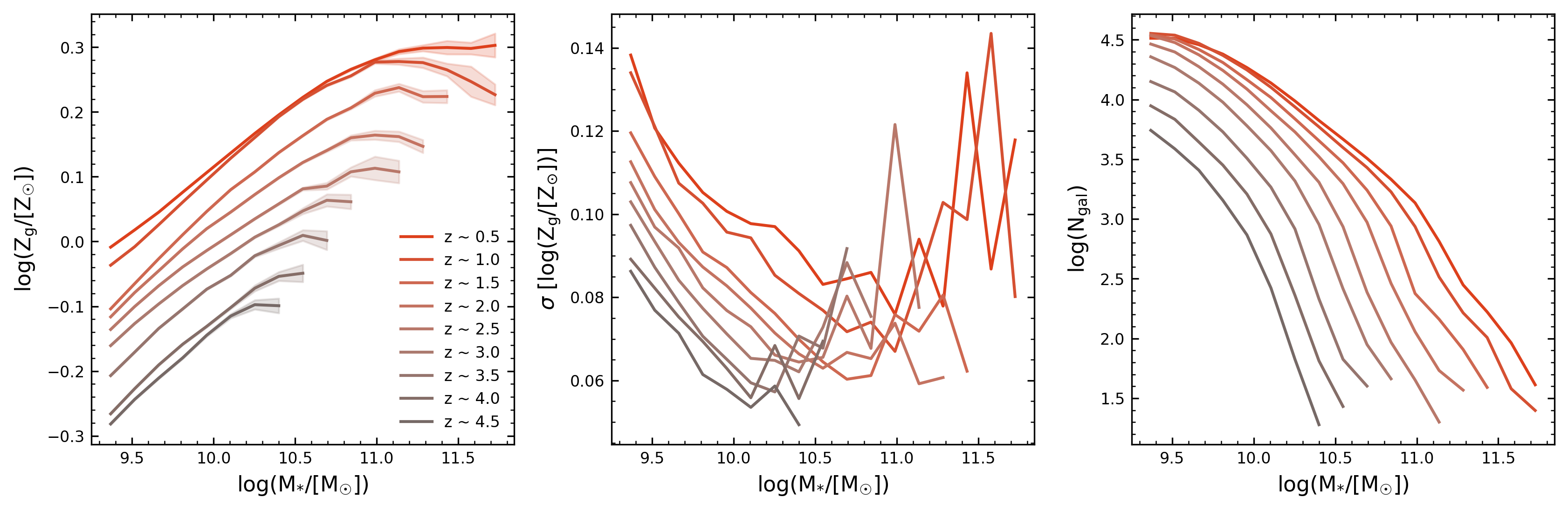}
    \caption{Left Panel: The MZR in HR5 for 9 snapshots from $z = 0.5$ to $4.5$ calculated as the median $\mathrm{Z_gas}$ in 15 consecutive, equal size bins in $M_{\star}$. Middle Panel: The standard deviation, $\sigma$, of the MZR in 20 consecutive, equal-size bins in stellar mass for the same 9 redshift snapshots. Right Panel: The Stellar Mass distribution, in terms of number of galaxies, for the 9 redshift snapshots. The different shades between grey and red represent different redshift snapshots, with redder being a more present snapshot. The shaded regions show the standard error on the median value in each bin.}
    \label{fig:MZR_Redshifts}
\end{figure*}

To measure the scatter in the MZR, we calculate the MZR residual, $\mathrm{d}{\log(Z_{\rm g}/Z_{\odot})}$. The residual is calculated as 

\begin{equation} 
\mathrm{d}\log(Z_{g}/Z_{\sun}) = \log\big( \, Z_{g}/Z_{\sun}\big) - \log\langle Z_{MZR}/Z_{\sun} \rangle \,
\end{equation}

\noindent where $Z_{\sun}$= 0.0134 corresponds to the Solar metallicity from \citet{asplund2009}, and $\langle Z_{MZR} / Z_{\sun} \rangle$ represents the MZR value from a linear regression fit to the linear regime of the total MZR in HR5. We define these values by fitting this linear regime to the MZR between the mass-resolution limit at $M_{\star} = 
2\times10^{9}\,\mathrm{M_{\sun}}$ and $1\times\,10^{10.4}\, \mathrm{M_{\sun}}$, the point at which the MZR begins to deviate from a linear trend \citep{Tremonti2004, Zahid2014}. For each galaxy, taking the difference between its $Z_{g}/Z_{\sun}$ value and the $\langle Z_{g}/Z_{\sun} \rangle$ value corresponding to its stellar mass provides an individual $\rm d\log(Z_{g}/Z_{\sun})$ value for the galaxy, describing how scattered this galaxy is from the average MZR. To account for the redshift evolution of the MZR in our study and how this affects our $\rm d\log(Z_{g}/Z_{\sun})$ values, we re-fit the MZR with another linear regression in each snapshot and calculate each residual relative to the MZR at different redshifts. In observations, and other simulations, the normalisation of the MZR is seen to fall whilst the slope of the low-mass regime is seen to be invariant, \citep{2008Maiolino, Torrey2019, Sanders2021}. This is correctly reproduced in HR5, \citep{Lee2021}, as seen in Fig. \ref{fig:MZR_Redshifts} and Fig. \ref{fig:MZR_Redshifts_Observations}. As such the slopes of our fits have a reasonably consistent value between $z=0.625$ to $z=5$, whilst their normalisation has a clear evolution to lower values over the same range.

Finally, to measure the evolution of a certain galaxy property, $X$, we calculate its rate of change, $\Delta X/\Delta t$, between two consecutive snapshots in look-back time $t_1$ and $t_{2}$, where $t_{2}$ is one snapshot further back in time than $t_{1}$. This quantity is defined as follows:

\begin{equation}
\frac{\Delta X}{\Delta t} = \frac{X_{1}-X_{2}}{t_{2}-t_{1}} \frac{1}{X_{2}} \times 100
\label{eq:dXdt}
\end{equation}

This equation is written such that the sign of the output is influenced only by the change in the property $X$ i.e. in look-back time, $t_2$ is always greater than $t_1$. This calculation is done for the total gas fraction in the galaxy, $f_{\text{gas}}$, and its gas metallicity, $Z_{g}$. These values represent the relative, instantaneous, percentage changes in the property per $\mathrm{Gyr^{-1}}$, compared to the previous snapshot. Note that in this work gas fraction is calculated as the ratio of gas-mass to total-mass within an individual galaxy, i.e. $f_{\mathrm{gas}} = M_{\mathrm{gas}}/M_{\mathrm{tot}}$.

\section{Results}
\label{sec:Results}

In this section, we present our results on how the MZR, the scatter in MZR and other galaxy properties evolve with redshift in the context of cosmic environments.

\subsection{Gas-phase MZR redshift evolution}

As the Universe evolves, the average gas metallicity and the stellar mass distribution of the galaxy changes. Whether the MZR maintains the same slope or if the slope changes with the universe's evolution has been studied in the literature \citep{2008Maiolino, Zahid2014}.
The left panel in Fig. \ref{fig:MZR_Redshifts} demonstrates the redshift evolution of the MZR in HR5. Note that this work assumes the Solar abundances of \citet{asplund2009}. From the figure, it is clear that HR5 reports a strong evolution of the average gas metallicities, with higher mass, higher metallicity galaxies emerging at later times; this is expected and is seen in both observations \citep{Lian2018} and other simulations \citep{Ma2016}. The turnover in the MZR also begins to emerge at $z = 3$ and below, for stellar masses $\log( M_{\star} / \text{M}_{\sun} ) \gtrsim 10.4$. We also only begin to see a significant number of galaxies with $\log( M_{\star} / \text{M}_{\sun} ) \gtrsim 11$ at $z \lesssim 2.5$. In the middle panel, we observe a clear negative correlation between the standard deviation of the total MZR, $\sigma_{\log(Z/Z_{\sun})}$, and $M_{\star}$ for low-mass galaxies, with the upper mass limit of the $\sigma$-$M_{\star}$ inverse correlation extending to consider galaxies with increasingly larger $M_{\star}$ at later cosmic times, as galaxies with intermediate values of $M_{\star}$ form in larger numbers as a function of time. This can be partially explained by the fact that galaxies with lower $M_{\star}$ are more sensitive to environmental effects, leading to a larger $\sigma_{\log(Z/Z_{\sun})}$ in the simulation. This scatter may also be regulated by mergers as a result of the central-limit theorem, as comparatively seen for the black hole mass ($M_{BH}$)-$M_\star$ relation in \citep{Hirschmann2010} that also shows reduced scatter for high stellar masses. A population of galaxies that under-goes a series of random mergers, not only pushes the population to higher-stellar masses, but also reduces the scatter in the $M_{BH}$-$M_\star$ relation, moving towards a normal distribution irrelevant of the correlation between the initial distributions of each variable, \citep{Zhu2025}. The relationship between $\sigma_{\log(Z/Z_{\sun})}$ and $M_{\star}$ breaks down in the high-mass end of the $M_{\star}$ distribution, which suggests a more complex evolution of the average gas metallicity of very massive galaxies, likely depending on specific environmental conditions. Lastly, in the right panel of Fig.\ref{fig:MZR_Redshifts}, we show the number of galaxies per stellar mass bin for each redshift, demonstrating the statistical significance of each snapshot in this space.


\begin{figure*}
    \centering
    \includegraphics[scale=0.65]{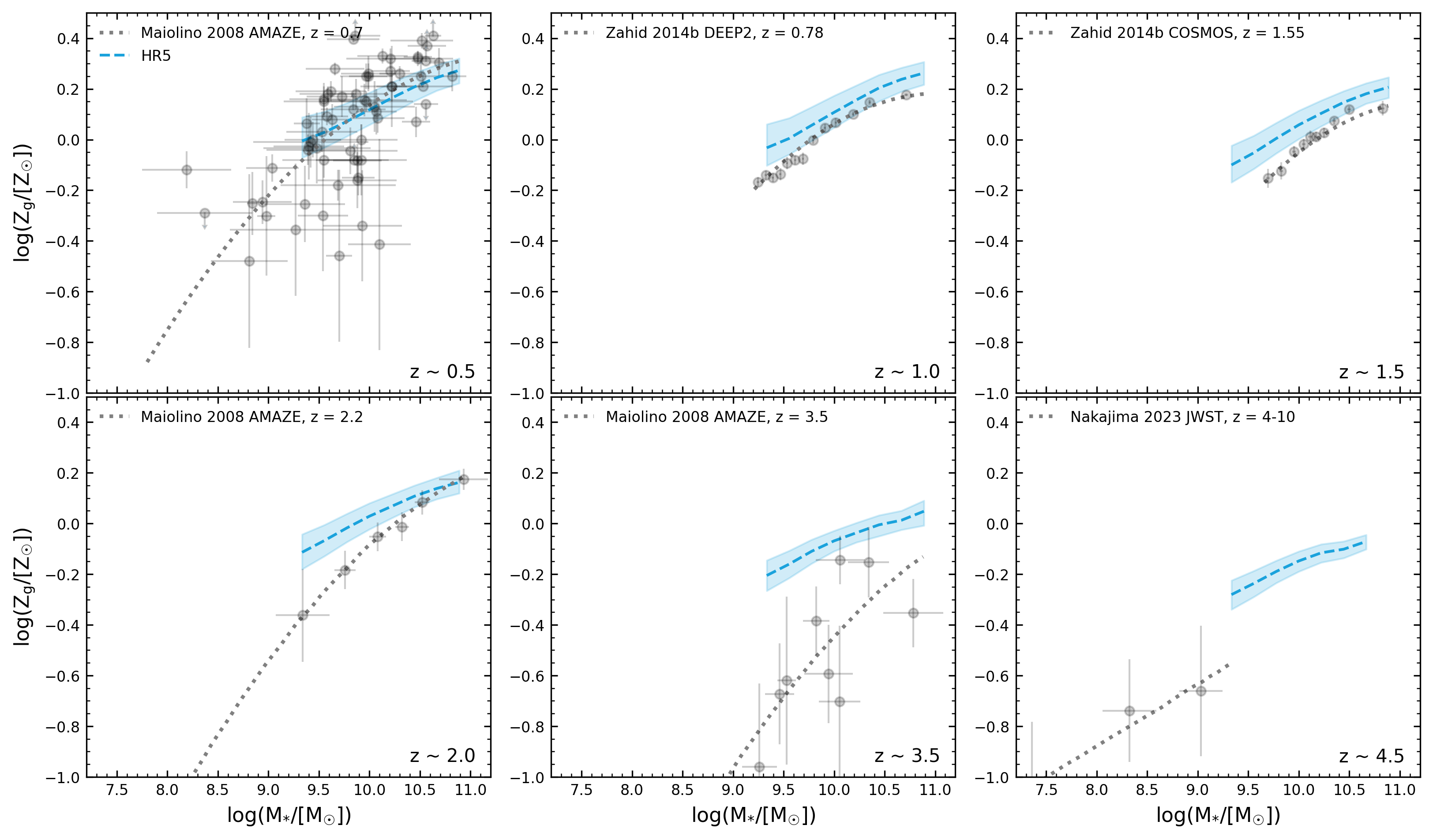}
    \caption{The total MZR in HR5 for 6 snapshots at redshifts, z, 0.5, 1, 1.5, 2, 3.5 and 4.5. The blue line shows the median HR5 MZR across 10 equal sized bins at the redshift specified in the bottom right corner of each subplot. The blue shaded region shows the interquartile range of the distribution in of metallicities in each bin. The dotted grey line shows the overall fit taken from the observed data points shown as the grey points from the study specified in the top left of each subplot.}
    \label{fig:MZR_Redshifts_Observations}
\end{figure*}

As mentioned in the methodology, one must be careful when comparing between observations and HR5 metallicities especially. HR5, and specifically its star formation history, is not well matched to observations to a very high accuracy, particularly at intermediate redshifts, $z=1$ to $3$. As seen in the original HR5 paper, \citep{Lee2021}, a comparison between HR5's cosmic star formation history (CSFH) and observations, \citep{Hopkins2004, Behroozi2013, Madau2014}, shows agreement at low, and high redshift, however a discrepancy between $z=1$ and $3$. Due to this, results referring to redshifts and mass dependences are purely relative within HR5's scales. To provide context to how HR5's MZR across redshift compares to observations, we include Fig. \ref{fig:MZR_Redshifts_Observations}. 
Each subplot contains the median HR5 MZR across our full stellar mass range in 10 consecutive equal sized bins, over plotted with an observed MZR taken from 1 of 4 studies. At z = 0.5, 2 and 3.5, we use the measurements given by \cite{2008Maiolino}, plotting the fitted line, and the points used to do so. At z = 1 and 1.5, we use \cite{Zahid2014} and their reported measurements from the DEEP2, at z = 0.78, and COSMOS, at z = 1.55, surveys. Finally at z = 4.5, we include more recent measurements from \cite{Nakajima2023}. It is evident that for lower redshifts HR5 is more agreeable with observations, particularly at its lowest snapshot at z = 0.5. For all redshifts HR5 either agrees or over predicts in $log(Zg)$, the magnitude of this over prediction increases with each consecutive snapshot. There are two ways to interpret this, either from the perspective of further observational considerations or further considerations in HR5. From the perspective of the simulation, the original Horizon Run 5 release paper, \citep{Lee2020}, shines light on this issue. It is suggested that the over-estimation may occur due to incomplete stellar feedback, resolution effects, and certain processes not being well captured and reproduced in the simulation. For in depth overview of these models, refer to \cite{Lee2021}. Outflows from galaxies typically host higher metallicities than the ISM average, \citep{Stinson2012}. These outflows can occur on sub-kpc scales, likely suggesting that they are not correctly accounted for in kpc scale simulations like HR5, which contributes to an over-estimation of metallicities compared to observations. From the alternate perspective, it is important to consider the possible selection bias in each observational sample. This bias will vary as a function of both redshift and stellar mass which impacts both the slope and gas-phase metallicity values of the sample. It is also important to mention the resampling method used in \cite{Zahid2014} This study uses bootstrap resampling to estimate the error on their median values. This method typically leads to underestimated error bars and as such more scatter around their results could be expected. In total, HR5s MZR is in a reasonable agreement with observations at a lower redshift when considering the caveats. As redshift increases, disagreement with observations does emerge.

\begin{figure*}
    \centering
    \includegraphics[scale=0.65]{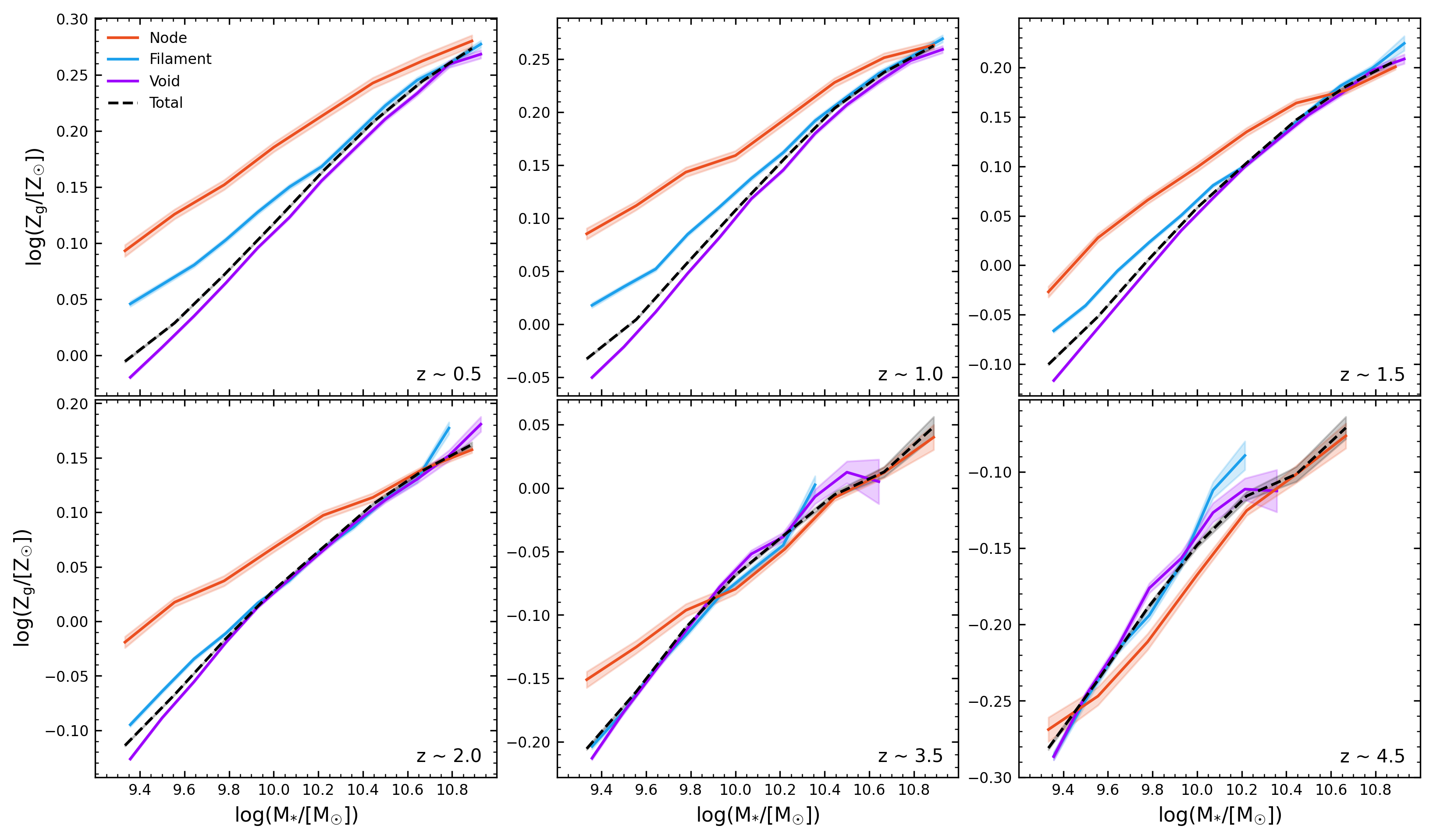}
    \caption{The MZR for node galaxies (orange), filament galaxies (blue), void galaxies (purple) and the total population (black dashed), over 10 consecutive, equal size bins in $M_{\star}$, for 6 of the chosen redshift snapshots. Bins with less than 15 galaxies have been removed. The shaded regions represent the standard error on the median value in each bin.}
    \label{fig:MZR_Scatter_Redshifts}
\end{figure*}

In Fig. \ref{fig:MZR_Scatter_Redshifts}, we show how the deviations from the total MZR emerge with redshift in the context of the three specified cosmic environments. We observe very low deviations from the MZR due to environment at $z = 4.5$, a time at which the three environments are indistinguishable from one another, agreeing with the result in Fig. \ref{fig:MZR_Redshifts}. As expected due to our previous work in \citep{Rowntree2024} and other existing studies, at $z = 0.5$, we observe the deviations due to the environment to be at their peak, with a noticeable difference between all environments. The redshift evolution shows that the offset from the MZR due to the node environment at low stellar mass emerges first; by $z = 4.5$ we already observe a $0.05\,\mathrm{dex}$ chemical enrichment for the node galaxies in the lowest stellar mass bins, with the filament and void galaxies in the same $M_{\star}$ range showing no signs of an offset in this snapshot. While only small deviations appear for filaments and voids at $z = 2$, the node galaxies now show a significant enrichment, as large as $\approx 0.1 \, \mathrm{dex}$ in the lowest $M_{\star}$ bin. Galaxies with higher $M_{\star}$ in the range $10^{9.8}$ - $10^{10.3}\,\text{M}_{\sun}$ also begin to show enrichment in the nodes at $z = 2$. For the lower $z$ snapshots, we observe that the three environments become distinct from each other, with nodes that remain highly enriched, filaments that are slightly enriched and voids that are under enriched relative to the total MZR in a particular snapshot. The upper limit on the $M_{\star}$ range enriched in nodes also continues to gradually increase snapshot by snapshot, with the full $M_{\star}$ range showing enrichment by $z = 0.5$. At this redshift, the low $M_{\star}$ galaxies again show the largest deviation from the total MZR. We see minimal change in the average gas metallicity of these galaxies between $z = 1$ and $0.5$, which seems to point to the presence of a ceiling on the metallicity offset due to the node environment for a fixed $M_{\star}$. For the highest $M_{\star}$ bins at all redshifts, the three environments are consistently indistinguishable from each other, with minor differences appearing at the most present snapshot.

\begin{figure*}
    \centering
    \includegraphics[scale=0.65]{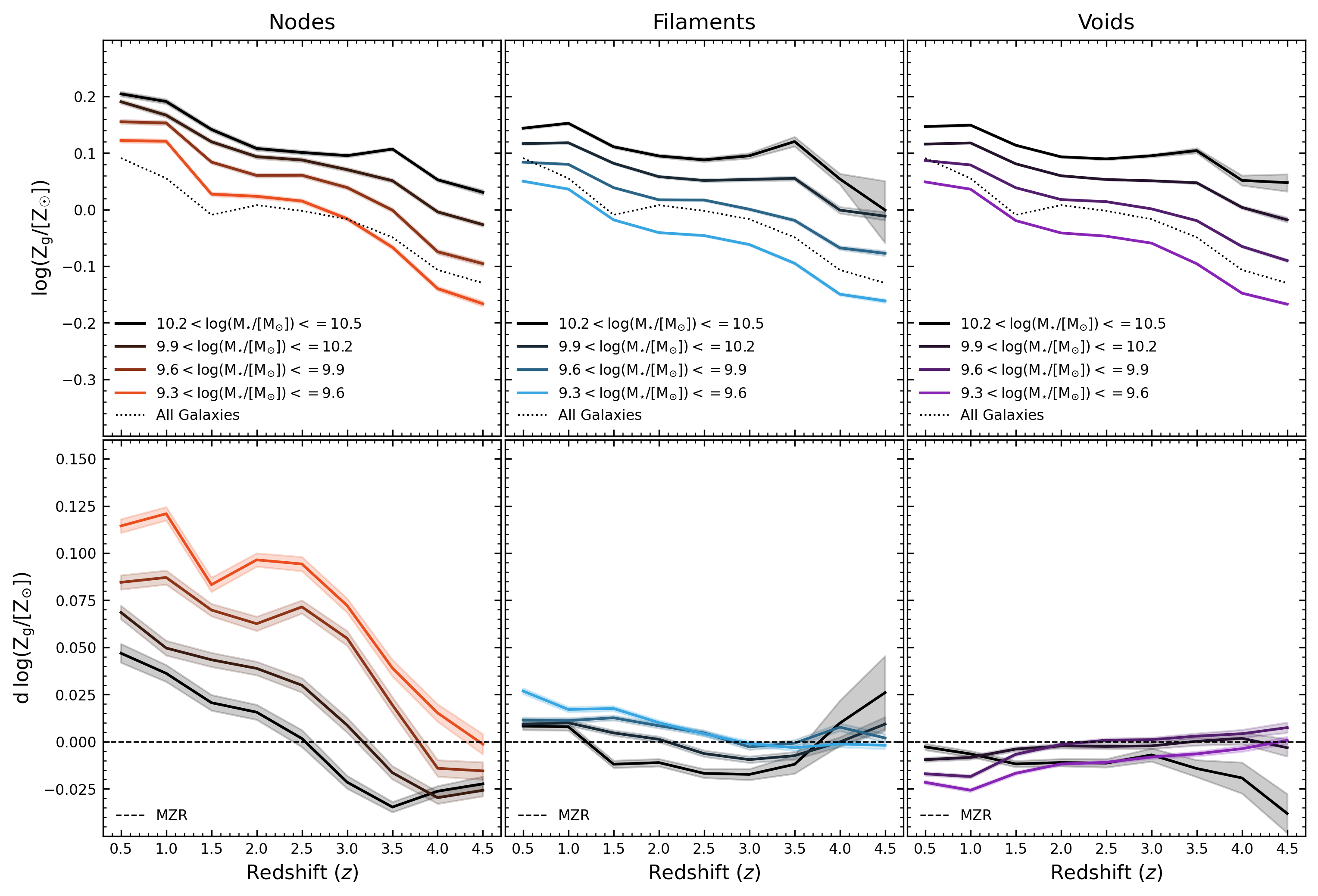}
    \caption{The redshift evolution of the gas metallicity (top row) and MZR residual (bottom row) of galaxies in nodes (orange), filaments, (blue) and voids (purple), in 4 equally spaced $M_{\star}$ bins (colour gradient). The shaded regions represent the standard error on the median. The dotted line in the top row represents the evolution of gas metallicity for the total population of galaxies. The dashed line in the bottom row shows the MZR value at each snapshot.} 
    \label{fig:Residual_Zgas_Redshift}
\end{figure*}

To further analyse how the deviations from the MZR and $Z_{g}$ evolve, we look to the residual $d\log(Z_{g}/Z_{\sun})$. Fig. \ref{fig:Residual_Zgas_Redshift} shows how the average $Z_{g}$ and $d\log(Z_{g}/Z_{\sun})$ vary against the redshift, $z$, between the three environments, in 4 distinct bins of stellar mass. In the top row, we see that the average gas metallicity increases with time, regardless of the environment, a simple outcome of our understanding of the Universe. However, the slope of this relationship is affected by the specific cosmic environment in question. In the nodes, we see a noticeably steeper relationship, with galaxies in this environment becoming enriched at a higher rate than those in filaments and voids, which are both similar in slope. In the literature these steepened trends in the nodes are are already accredited to shorter processing times in these dense environments, \citep{Peng2010, Wetzel2013, Ditrani2025} and our finding aligns with this picture. As expected, more massive galaxies exhibit higher metallicities; this persists in all environments and at all redshifts. In the bottom row, we look at how MZR residuals change with redshift. The nodes show an increasing $d\log(Z_{g}/Z_{\sun})$ with decreasing redshift. The trend has the same direction in the filaments, however the opposite is true for the voids within which $d\log(Z_{g}/Z_{\sun})$ decreases as time passes, As expected, the evolution of the residual is accelerated in the nodes, reaching larger magnitudes for the same epoch as the filament and void residuals. All trends show a strong stellar mass dependency with the low mass galaxy population showing the largest residuals for nearly all snapshots. Interestingly, the direct difference between the voids and the other two environments is clear here; low stellar mass galaxies in voids still show the largest magnitude of residual, however the evolution with stellar mass is towards negative values rather than positive. By $z = 0.5$ the low mass population in the nodes has reached a maximum residual of $0.13$ dex, whilst the same mass bin in filaments and voids show, $0.04$ dex and $-0.03$ dex respectively.

\subsection{Radial profiles of clusters and filaments}

\begin{figure*}
    \centering
    \includegraphics[scale=0.65]{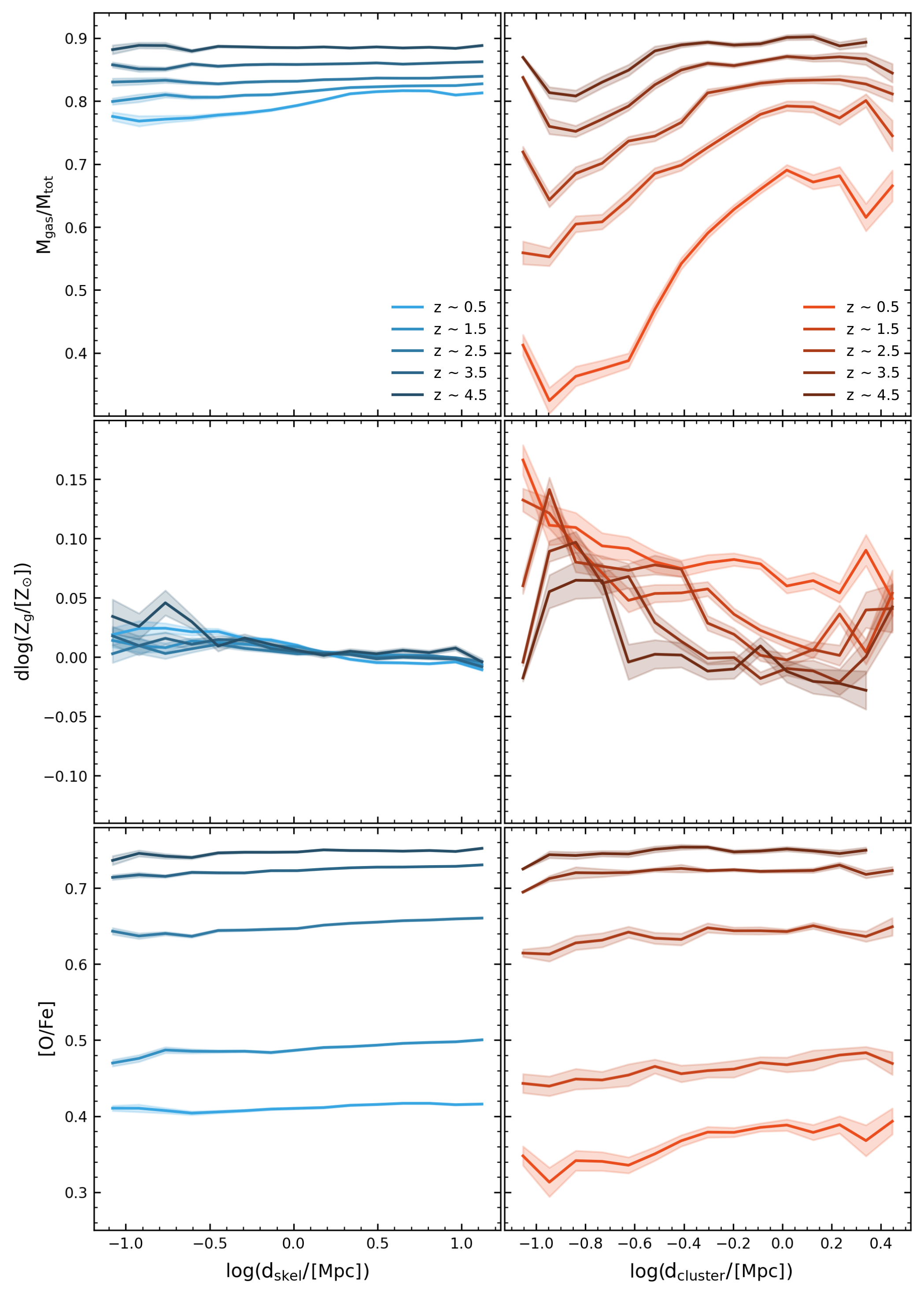}
    \caption{Radial profiles around filaments (blue) and nodes/clusters (orange) for gas fraction (top row), Residual from the MZR (middle row) and [O/Fe] (bottom row) and their evolution with redshift from $z = 0.5$ to $4.5$ with brighter colours representing a more present time. The shaded regions show the standard error on the median value in each bin.}
    \label{fig:Radial_Profiles_dskel_dcluster}
\end{figure*}

Fig. \ref{fig:Radial_Profiles_dskel_dcluster} presents radial dependencies of gas fraction, $f_{\text{gas}}$, MZR residual, $d{\log(Z_{g}/Z_{\sun})}$, and $\mathrm{[O/Fe]}$ around filaments, $d_{\rm skel}$, and nodes/clusters, $d_{\rm cluster}$, across the full redshift range. In the top row, we see a clear drop in gas fraction at low redshifts around the core/centres of both structures; this effect is dramatically stronger in nodes and stronger at lower redshifts, with a consistent evolution where each snapshot shows a larger decrease in $f_{\text{gas}}$ with radius than the last. 
The middle row shows the relationship between $d{\log(Z_{g}/Z_{\sun})}$ with $d_{\rm skel}$ and $d_{\rm cluster}$. In the first column, we see how $d{\log(Z_{g}/Z_{\sun})}$ varies with $d_{\rm skel}$, for all redshifts, this trend is very weak. The largest increase in residual, $0.04$ dex, from $10$ to $0$ cMpc does exist in the latest snapshot, $z = 0.5$, with higher redshifts showing a flatter trend; however, it is difficult to make any concrete conclusions from such a weak trend. This is the opposite in the Nodes, where the trends are far more pronounced. Interestingly, in the centre of clusters, $d_{\rm cluster} < -0.6$ log[cMpc], for all redshifts, chemical enrichment exists, with slightly higher residuals existing for consecutive lower redshift snapshots. At $z = 0.5$, a maximum residual of $0.17$ dex above the MZR exists in the central most bin. Generally, as the redshift increases, the maximum residual in each snapshot falls. At $z = 4.5$, the maximum residual has fallen to $0.07$ dex. Another difference between the snapshots in this space is that at high redshift ,$z = 4.5$, the cluster outskirts are not enriched relative to the background, the only thing observed is a large residual increase near the centers of the Nodes,  and a flat trend outside of this region. In this snapshot, an increase in residual of $\approx 0.09$ dex occurs but only within $d_{\rm cluster} < -0.6$ log[cMpc]. As time passes, this outskirts region increases in residual, creating a more consistent trend where residual increases with a falling $d_{\rm cluster}$. By $z = 0.5$, the trend has evolved to show a consistently increasing residual from $0.05$ dex at the very cluster outskirts to $0.17$ dex at the peak of the trend. This is a total increase in this snapshot of $\approx 0.12$ dex, which is similar to the increase seen at $z = 4.5$; however, it now occurs over the full range of $d_{\rm cluster}$, rather than just the inner cluster, due to the growth of the halos in time. The last row of Fig. \ref{fig:Radial_Profiles_dskel_dcluster} shows radial profiles around the structures for $\mathrm{[O/Fe]}$. The strongest trend observed for these sub-figures is the redshift evolution of $\mathrm{[O/Fe]}$. Lower $\mathrm{[O/Fe]}$ at later times is an expected result as Type Ia Supernovae (SNeIa) contribute more to the iron abundance as time goes on, these values also suggest relatively older stellar populations in these galaxies. Radial trends do emerge for both environments, again more strongly for the Nodes, leading to a $\approx 0.08 \, \mathrm{dex}$ change at $z = 0.5$ for $\log(d_{\rm cluster}/\text{cMpc})$ in the range from $0.4$ to $-1$, suggesting stronger environmental quenching effects in these regions. In filaments, the largest slopes seem to emerge for intermediate redshifts ($z = 1.5$); however, these weak relationships show a change of $\approx 0.03 \, \mathrm{dex}$ over the same distance range.

\subsection{Evolution of [O/Fe]}
\begin{figure*}
    \centering
    \includegraphics[scale=0.65]{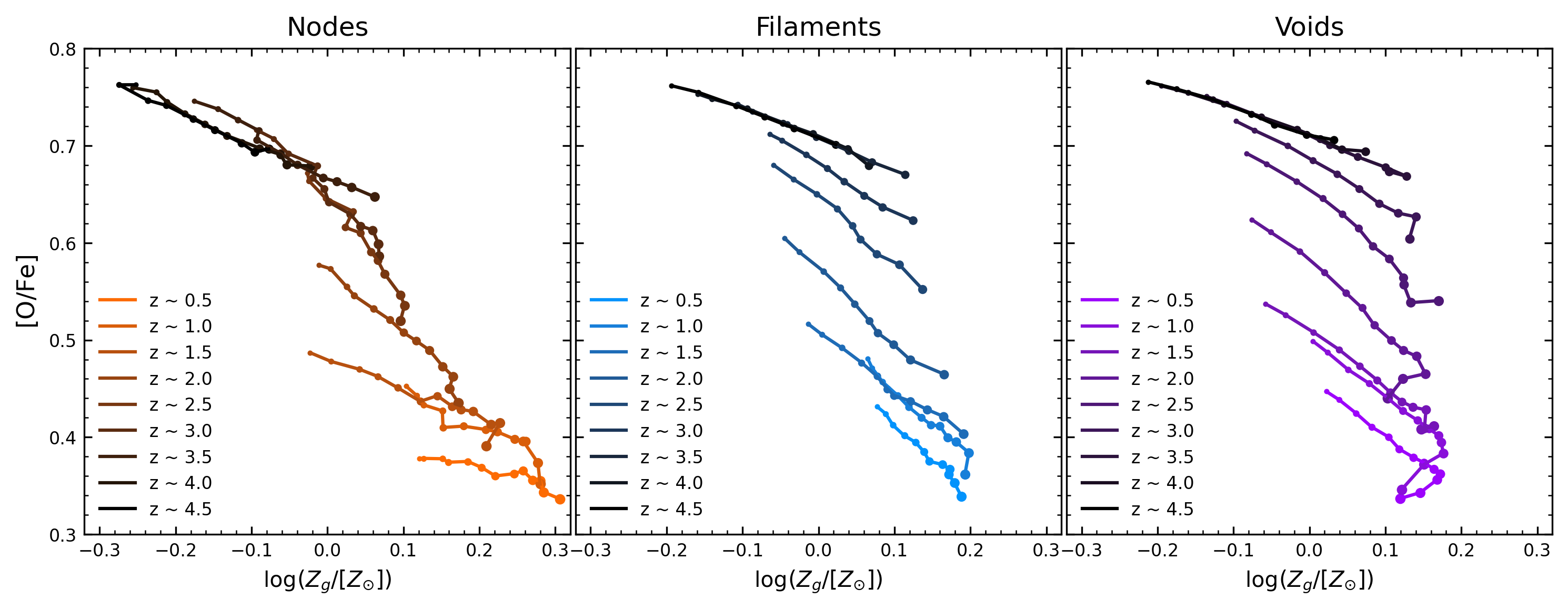}
    \caption{The median gas metallicity, $Z_{g}$, of galaxies in nodes (orange), filaments (blue) and voids (purple), against the median [O/Fe] value in 15 equal-sized bins in [O/Fe] between the minimum and maximum values in [O/Fe] at each redshift. The size of the points on each line represents the median stellar mass of the population in the bin, demonstrating relative increases or decreases in this quantity, whilst brighter colours demonstrate a lower redshift as shown by the legend.}
    \label{fig:OFe_LogZ}
\end{figure*}

We also examine how [O/Fe] varies between environments in Fig. \ref{fig:OFe_LogZ} in a different plane. Here, we present the evolution of [O/Fe] against $\mathrm{Z_{gas}}$ in redshift; the filled circles display the median stellar mass of the galaxies populating each bin in $Z_{g}$. 
First it is important to acknowledge the enhanced values of [O/Fe] relative to observations, within similar redshift ranges, see \citep{Gibson1997,chiappini1997}. Observationally it is expected that for lookback times $< 13 Gyrs$, [O/Fe] sits in the range of $-0.1 < \mathrm{[O/Fe]} < 0.2$. HR5 reports significant enhancement relative to these numbers still reaching an [O/Fe] of 0.5 around $\mathrm{z \approx 2}$. This is a common occurrence in large cosmological-hydrodynamical simulations, seen also in IllustrisTNG and EAGLE \citep{Chruslinska2024}, although to a slightly lesser degree. HR5 over-estimates when compared to these other simulations by approximately 0.1-0.15 dex. \citep{Lee2021} shows the evolution of $\mathrm{12+log(O/H)}$ compared to observations, between $z\approx0.7$ and $3.5$, and clearly shows an significant offset in metallicity at both $z\approx3.5$ and $z\approx1.6$. This suggests that SFR in HR5 may not be well regulated, possibly due to weak feedback and more specifically an underestimation of the contribution from prompt SNeIa.

The Type 1a Sne delay time distribution (DTD) implemented in RAMSES-CH is one based on studies in \cite{Few2014} in which the variance in the DTD for a single disc galaxy was shown to significantly effect the slope, magnitude and shapes of trends in [O/Fe] against metallicity. Although this is an isolated example compared to the large dataset we use, the DTD analysis in this study directly impacted the model that is used in HR5. The DTD chosen was the most delayed model as more prompt models could not reproduce the observed "knee" in the [O/Fe] on [Fe/H] plane. As such the contribution from prompt type 1a SNe in RAMSES-CH, and therefore HR5, is low. Although HR5 overestimates the values for [O/Fe] compared to observations, and it is clear the choice of DTD effects the slope and magnitude of these values, only a single DTD is implemented in which our results are based on. As such, relative changes in our trends due to other factors, like environment, still retain their importance. Quantitative comparisons between the results in HR5 and observations are to be made carefully; our study aims to focus on relative changes within our isolated system due to the factors we seek to understand.

An additional enhancement of the contribution from prompt SNeIa in the model would likely produce ratios that are in more agreement with those seen in nature, without impacting upon the relative trends or conclusions from our analysis.
The first distinct feature in this figure between the environments is the difference in the evolution of the slope. Generally, a negative correlation exists with higher stellar mass and gas metallicity galaxies having lower [O/Fe] values. In the nodes, at both the highest and lowest redshift, the slope of the trend is at its flattest. From $z = 4.5$, within three snapshots the trend has steepened between $z = 3.5$ and $2$. For $z \lesssim 2$, the trend flattens once again, continuing to flatten until the most present snapshot at $z = 0.5$. This slope evolution is completely different in the filamentary environment; again, it begins flat at $z = 4.5$; however, there is now a consistent steepening down to $z = 0.5$, with no signs of flattening. This changes once again in the Voids; just as in the filaments, we do see a consistent steepening of the trend down to $z = 0.5$; however, at $z = 3$, we begin to see a turnover in [O/Fe] for the high stellar mass galaxies. This turnover persists in all redshift bins below $z = 2$, demonstrating that high stellar mass galaxies are losing metallicity whilst still experiencing a decreasing [O/Fe] when in the void regions.

\subsection{Rates of change of galaxy properties}

\begin{figure*}
    \centering
    \includegraphics[scale=0.65]{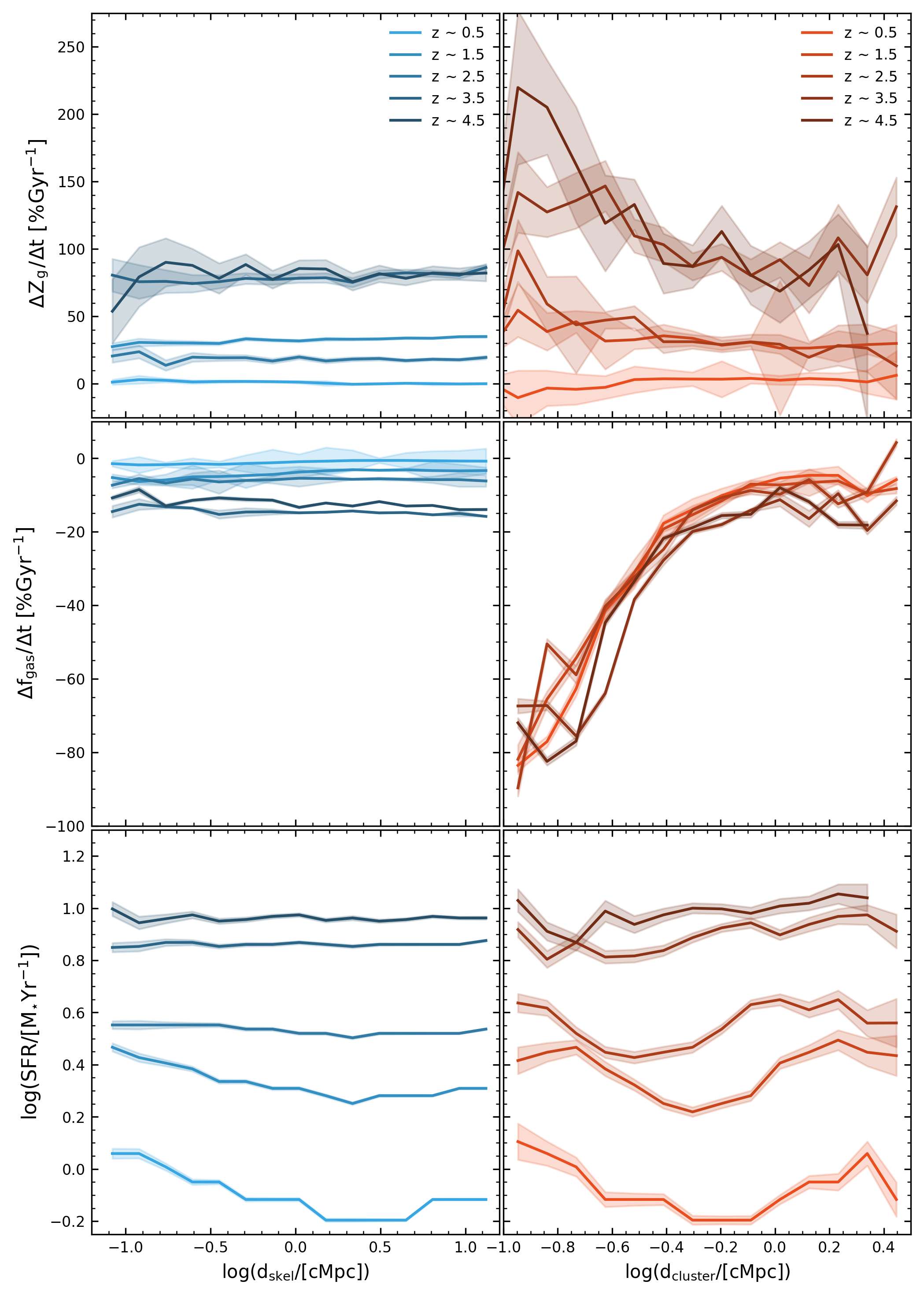}
    \caption{Radial profiles around filaments (blue) and nodes/clusters (orange) of the instantaneous evolution parameters $\Delta Z_{g}/\Delta t$ (first row), $\Delta f_{\text{gas}}/\Delta t$ (second row), $\log(\text{SFR})$ (third row), split into 15 equal-size logarithmic bins in $d_{\text{skel}}$ (left) and $d_{\text{cluster}}$ (right). The shaded region represents the standard error on the median. Evolution with redshift from $z = 0.5$ to $4.5$ is shown using the colour gradient, with brighter colours representing a more present time. Refer to equation \ref{eq:dXdt} for the calculation of the percentage rate of changes that are used.}
    \label{fig:dXdt_Radial_Profiles}
\end{figure*}

Fig. \ref{fig:dXdt_Radial_Profiles} introduces the $\Delta X/\Delta t$ evolution parameters, providing a measure of the rate at which certain galaxy properties, $X$, are changing at an instantaneous point in time. We provide the percentage rate of change in gas metallicity, $\Delta Z_{g}/\Delta t$, and in gas fraction, $\Delta f_{\text{gas}}/{{\Delta t}}$. We also include the star formation rate (SFR) in this part of the analysis. Fig. \ref{fig:dXdt_Radial_Profiles} shows radial profiles with $d_{\rm skel}$ and $d_{\rm cluster}$ for these evolution parameters. At first glance, it is clear that nodes present a much stronger relation in most parameters, with more variation in their median $\Delta Z_{g}/\Delta t$ between adjacent bins. There is already more variation in this property around both filaments and nodes, it is likely that the smaller bin populations surrounding the clusters enhance this effect. In the top row, we look at the radial profiles of $\Delta Z/{\Delta t}$. Around both structures, the highest redshift snapshot, $z = 4.5$, shows the highest values of $\Delta Z/{\Delta t}$ across the full distance range. With $d_{\rm cluster}$, $\Delta Z/{\Delta t}$ shows significant variance at these high redshifts where we observe that galaxies in the centres of clusters show a rapidly increasing metallicity when compared to galaxies in the cluster outskirts. The cluster outskirts values match the values seen in the filaments at the same redshift, suggesting that these regions are reflective of the background values of $\Delta Z/{\Delta t}$. When moving to lower redshifts, the trends around clusters begin to flatten, with galaxies slowing down their evolution of $Z_{g}$ and becoming more stable. By $z = 0.5$, $\Delta Z/{\Delta t}$ has settled around 0 for all values of $d_{\rm skel}$ and $d_{\rm cluster}$, and any existing trend with $d_{\rm cluster}$ has vanished. Note that, the median $\Delta  Z/\Delta t$ is positive, indicating an increasing gas metallicity with time at all snapshots. The middle row of Fig. \ref{fig:dXdt_Radial_Profiles} shows the radial profiles of $\Delta f_{\rm gas}/\Delta t$, which refers to the total gas fraction of galaxies. At all distances and redshifts, the median of this parameter is negative, stating a loss of gas. It is also important to note that the redshift evolution of this parameter across these distance ranges is small; in other words, the median change in the gas fraction of all galaxies is more similar than that of other parameters across all redshifts. In filaments, the radial trends are very weak with little to no change across the full distance range; however, it is obvious that this parameter is important for node environment. Within $\log( d_{\rm cluster}/\text{cMpc} ) \approx -0.2$, there is a dramatic drop in $\Delta f_{\text{gas}}/\Delta t$ at all redshifts of the same magnitude suggesting something fundamental about how galaxies are losing their gas in the centres of clusters. $\Delta f_{\text{gas}}/\Delta t$ reaches the median values seen in the filaments as we move further away from the nodes for $\log( d_{\rm cluster}/\text{cMpc} ) \gtrsim -0.2$ which are likely the background field values. In the bottom row of Fig. \ref{fig:dXdt_Radial_Profiles} we show the radial profiles of the SFR. Again, the trends with $d_{\rm skel}$ are reduced when compared to $d_{\rm cluster}$; however, they are relatively similar to other parameters. A slight trend exists at $z = 0.5$ for both filaments and nodes in their cores, showing a heightened SFR not present at the higher redshifts. Around the nodes, this heightened SFR in the core is present in more snapshots; however, rather than flattening the trend as seen in filaments, we observe a dip in SFR away from the core of the nodes. The distance at which the dip in SFR occurs increases from $ \log( d_{\rm cluster}/\text{cMpc} ) \approx -0.55$ at high redshift ($z = 3.5$) to $\log( d_{\rm cluster}/\text{cMpc} ) \approx -0.2$ at low redshift ($z = 0.5$), likely due to the growth of the halos with decreasing redshift. Interestingly, either side of this specific distance at a set snapshot shows an increase in SFR, suggesting a specific part of the cluster is more quenched than others. It is possible that this trend can be attributed to the presence of filaments that connect to the cluster at its outskirts. As galaxies within clusters tend to be distributed correlating with the filament connections, the increased SFR in the cores of filaments could permeate into the node environment, showing the same magnitude of increased SFR at their edges.

In Fig.\ref{fig:Radial_Profiles_dskel_dcluster}, at these larger distances $d_{\rm cluster}$, we also observe a downturn in the gas fraction corresponding to this increase in SFR, which could imply that at the cluster outskirts, gas is being consumed to produce stars, increasing SFR to levels also observed in the cores of these structures. To truly probe this effect, one would need to look specifically at the cold gas fraction, rather than the total gas fraction which we use in this study. The variance in the SFR appears between the cluster outskirts and the core rather than a simple negative correlation with $d_{\rm cluster}$, which one may intuitively expect.

\begin{figure*}
    \centering
    \includegraphics[scale=0.65]{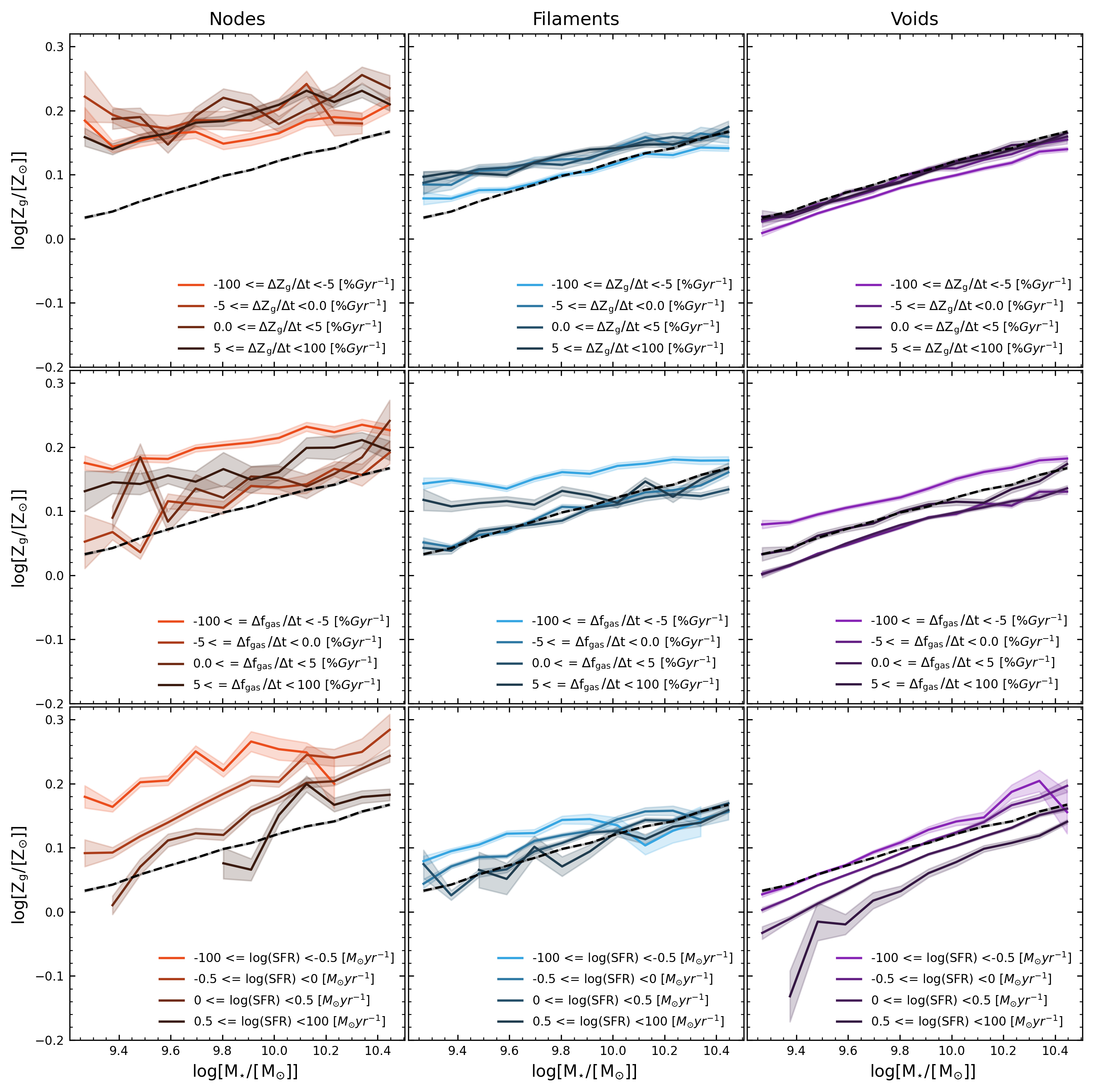}
    \caption{Individual MZRs for each environment, nodes/clusters (orange), filaments (blue) and voids (purple) binned by each instantaneous evolution parameter, demonstrating the deviation from the MZR due to cosmic environment. $\Delta Z_{g}/\Delta t$ (first row), $\Delta  f_{\text{gas}}/\Delta t$ (second row), $\log(\text{SFR})$ (third row) split into 15 equal-size logarithmic bins. The shaded regions show the standard error on the median. The saturation of the lines represents different bins in each parameter, with the more colourful lines representing a strong negative evolution and the darker lines showing a strong positive evolution. The dashed black lines show the trend for the total population of galaxies. The plot is carried out only for the most present snapshot at $z = 0.5$. Refer to equation \ref{eq:dXdt} for the calculation of the percentage rate of changes that are used.}
    \label{fig:MZR_Scatter_dXdt}
\end{figure*}

To see how these evolution parameters contribute to the scatter in the MZR, we split the population of galaxies in each environment into four separate bins of each parameter. The binning is done such that we have a strongly negative bin, weak negative bin, weak positive bin and strong positive bin to represent four cases where a galaxy could be evolving in each parameter. We then plot a specific MZR for each bin and compare them. 
Fig. \ref{fig:MZR_Scatter_dXdt} shows the scatter due to these evolution parameters in the last snapshot of HR5 at $ z = 0.5$. In the top row, we consider $\Delta Z_{g}/\Delta t$ there is no consistent scatter due to this parameter; however, for filaments and voids, we do see galaxies that have a strong negative $\Delta  Z/\Delta t$ to have noticeably lower gas metallicities than the other bins. A clear trend that does exist is the increase in variance between all bins from voids to filaments to nodes. In voids there is minimal variance in the MZR for each of the $\Delta  Z/\Delta t$ bins, sitting at a value of $ \approx \, 0.01 \mathrm{dex}$, in filaments this variance has slightly increased to $ \approx 0.03 \, \mathrm{dex}$, and in node environment has significantly increased to $ \approx 0.09 \, \mathrm{dex}$. \par 

In the middle row of Fig. \ref{fig:MZR_Scatter_dXdt} we explore the effect of $\Delta f_{\text{gas}}/\Delta t$. Here, we can see a more consistent scatter emerge; galaxies that are losing or consuming gas fast in all environments across all stellar masses are enriched relative to the MZR at their specific stellar mass. The galaxies with the lowest gas metallicities over the full stellar mass range do not have a rapidly changing gas fraction as we also see a rapid increase in gas fraction also leads to enrichment compared to the two low magnitudes $\Delta f_{\text{gas}}/\Delta t$ bins. These relationships are most clear in filaments and voids but still exist in the nodes. Note that in filaments and marginally in the most negative bin in nodes, we see a slight tendency for low stellar mass galaxies to have higher gas metallicities in this scatter. The last row of Fig. \ref{fig:MZR_Scatter_dXdt} repeats these cuts for SFR. A stark difference is clear for the three environments, with a large consistent scatter emerging in nodes, $\approx 0.2 \, \mathrm{dex}$, very little scatter in filaments $\approx 0.03 \, \mathrm{dex}$, and again a clear scatter in voids, $\approx 0.1 \, \mathrm{dex}$. As we observed when investigating the environmental effect on the fundamental metallicity relation in \citet{Rowntree2024}, it is clear that when binned by SFR, the slope of the MZR for each environment is different. The nodes and voids show a steeper slope compared to the total MZR slope, whilst the filaments show a slightly more shallow relationship.
Interestingly, in the nodes, the trends show that galaxies with higher stellar mass are more enriched and have higher residuals than their lower stellar mass counterparts when binned by SFR. This is the opposite for the other parameters and environments, where low stellar mass galaxies show the largest residuals. Note that the overall correlation is significantly reduced in the filament environment. In all environments across the full stellar mass range, galaxies with the lowest star formation rate show the highest metallicities. Since these galaxies also show the most negative $\Delta f_{\text{gas}}/\Delta t$, these could be objects that are rapidly quenching. This trend persists for higher SFR bins, where each has a relatively low average gas metallicity when compared to the lower SFR case.

\begin{figure*}
    \centering
    \includegraphics[scale=0.65]{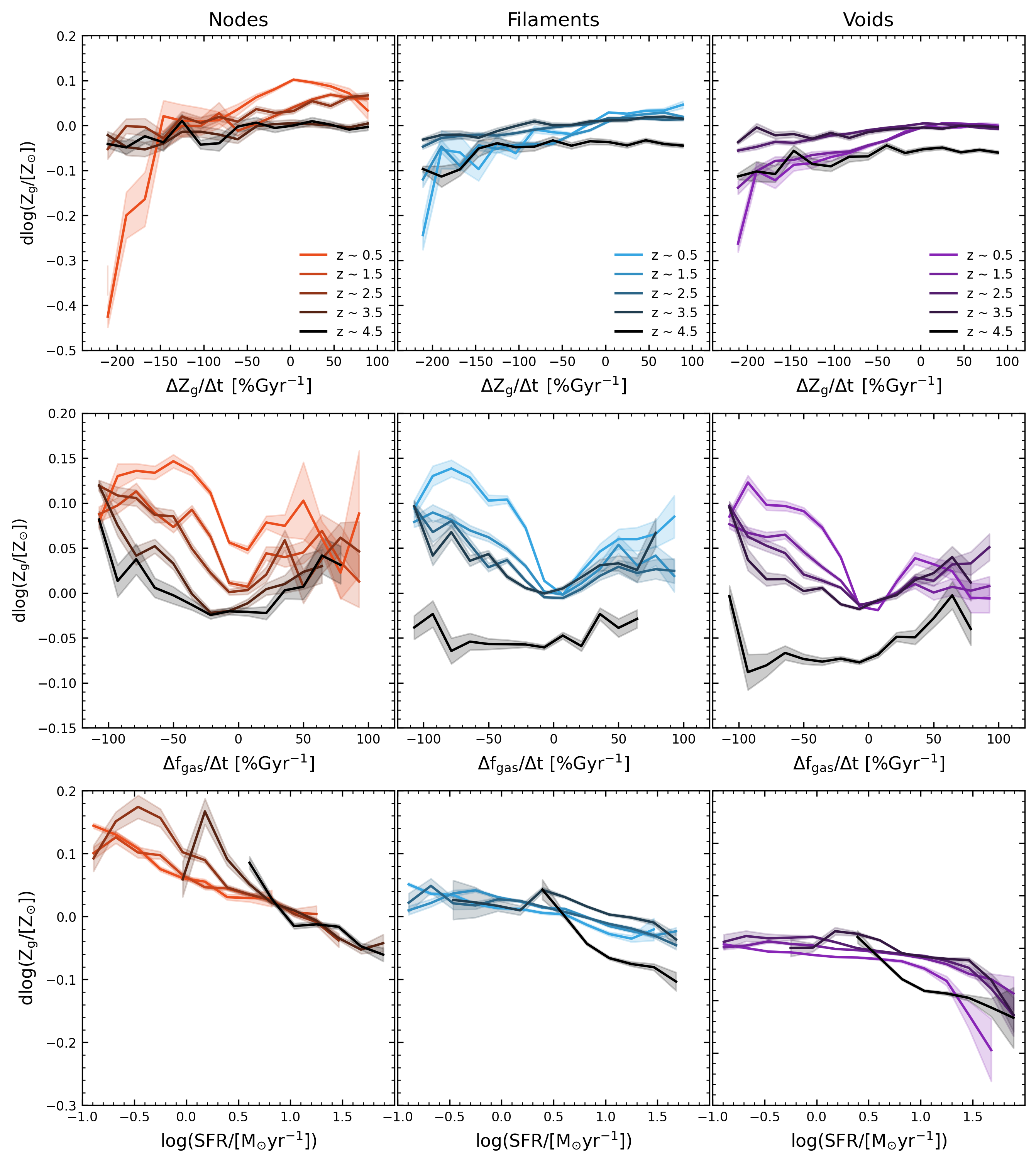}
    \caption{Each $\Delta X/\Delta t$ parameter and its relationship with the residual, $d\log(Z_{g}/Z_{\sun})$, from the total MZR within each environment, nodes (orange), filaments (blue) and voids (purple), at 5 snapshots from $z = 4.5$ to $0.5$. Each line represents the median value of $d\log(Z_{g}/Z_{\sun})$ in 15 equal size bins in the property in question. The shaded region shows the standard error on this median. Each sub-figure is coloured such that the brightest colour represents the lowest redshift snapshot, and vice versa for the highest redshift snapshot as shown in the legend. Refer to equation \ref{eq:dXdt} for the calculation of the percentage rate of changes that are used. }
    \label{fig:dXdt_Residuals}
\end{figure*}

For further insight into how these evolution parameters affect the MZR and its scatter, we now look at Fig. \ref{fig:dXdt_Residuals}, which shows their relationships with the residual, $d{\log (Z_{g}/Z_{\sun})}$, for each of the snapshots and each cosmic environment. In the first row, the trend with $\Delta Z_{g}/\Delta t$ ties together the results shown in Fig. \ref{fig:dXdt_Radial_Profiles}. In this space, positive values of $d\log(Z_{g}/Z_{\sun})$ only occur above $\Delta Z_{g}/\Delta t \approx  -50$ for low redshifts, $z < 2.5$. With the highest residuals, $d\log(Z_{g}/Z_{\sun}) \approx 0.1$ dex, only appears for $\Delta Z_{g}/\Delta t > 0$. Interestingly, as a function of distance to cluster of filament centre, higher values of $\Delta Z_{g}/\Delta t$ only occur for in the centres of the clusters, with filaments not mimicking this increased rate. In Fig. \ref{fig:dXdt_Residuals} all three environments show similar trends, however filaments and voids do not display a larger positive residual for $\Delta Z_{g}/\Delta t > 0$ that the node population does. For galaxies rapidly dropping in $Z_{g}$, $\Delta Z_{g}/\Delta t < -150$, a significant drop in $d\log(Z_{g})$ is observed with varying magnitudes based on the environment. Nodes show a decrease of $\approx 0.42$ dex between $-200 < \Delta Z_{g}/\Delta t < -150$ whilst filaments and voids both show a $\approx 0.17$ dex decrease. This drop only occurs in the latest snapshot, $z = 0.5$, although all other snapshots show a weaker positive correlation between the two properties, which flattens as you move to higher redshifts. \par 

The middle row of  Fig. \ref{fig:dXdt_Residuals} shows how $d\log(Z_{g}/Z_{\sun})$ is affected by $\Delta  f_{\text{gas}}/\Delta t$. Broadly, a galaxy that is gaining or losing gas shows an increased $d\log(Z_{g}/Z_{\sun})$, demonstrated by the large dip in residual at $\Delta  f_{\text{gas}}/\Delta t = 0$. Within this trend, galaxies that are rapidly losing gas, $\Delta  f_{\text{gas}}/\Delta t < -50$, show the highest positive residuals for each snapshot. Galaxies with $\Delta  f_{\text{gas}}/\Delta t < -50$ at $z = 0.5$ show a maximum residual of $\approx 0.15$ dex. Galaxies that are gaining gas also show an increased residual, however, to a lower magnitude than that seen for gas loss. In the same snapshot, galaxies with $\Delta  f_{\text{gas}}/\Delta t > 50$ show a maximum residual of $\approx 0.08$ dex. The highest residuals are seen in the latest snapshot, $z = 0.5$, for all environments, where each consecutively earlier snapshot shows a reduced residual and a weakened trend across the full range of $\Delta  f_{\text{gas}}/\Delta t$. The overall trend persists between all 3 environments, the only difference being the expected result that filaments and voids show reduced residuals when compared to the nodes. Note that at the earliest snapshot, $z = 4.5$, there is a significant difference in the minium residual between the environments across the full range, with voids showing the lowest $d\log(Z_{g}/Z_{\sun})$ and nodes showing the highest $d\log(Z_{g}/Z_{\sun})$. \par 

The final row of Fig. \ref{fig:dXdt_Residuals}  exhibits the trend between the residual and SFR. As expected, it is seen in all environments that a lower SFR leads to a higher $d\log(Z_{g}/Z_{\sun})$, indicating that we are looking at passive galaxies in this range. The slope of this trend is dependent on the environment, with the nodes showing a steep trend, filaments intermediate, and voids almost flat. The highest residual values, $\approx 0.15$ dex, are found in the nodes for $\log(\text{SFR}) < -0.5$ dex for $z = 2.5$ where the statistical sample is sufficiently large to populate this region of the space. Note that in the voids, the trend is largely flat until high SFR, $\log(\text{SFR}) > 1$ dex is reached, after which a significant drop of $0.1$ dex is observed in all snapshots. \par 

These results suggest that the high positive residual population at $z = 0.5$, $d\log(Z_{g}/Z_{\sun}) > 0.1$ dex discussed throughout this section can be coupled together with high $\Delta Z_{g}/\Delta t$, low $\Delta  f_{\text{gas}}/\Delta t$ and low SFR and will likely exist within centers of nodes, as seen in Fig. \ref{fig:dXdt_Radial_Profiles}. On the other hand, the low residual population, $d\log(Z_{g}/Z_{\sun}) < -0.1$, is associated with very low $\Delta Z_{g}/\Delta t$, a $\Delta  f_{\text{gas}}/\Delta t$ of $\approx 0$ and a higher SFR.

\section{Discussion}
\label{sec:Discussion}

Our results suggest a spatially dependent chemical enrichment in the Universe across cosmic time, within which galaxies begin to show deviations from the MZR at approximately $z = 3.5$ for the most dense regions. Galaxies that have fallen into galaxy clusters experience an accelerated enrichment compared to galaxies in under-dense regions, possibly due to the intense processes that are at play in these regions, i.e. galaxy mergers, ram-pressure stripping, increased AGN-activity and various channels of starvation. 
\par

At low redshifts there is conflicting evidence to suggest whether nodes, or clusters, show reduced, or increased merger rates when compared to smaller galaxy groups in the field or isolated systems. \citep{Moss2006, Liu2009, Ellison2010} all present results that indicate increased merger fraction for galaxies in high-density environments at $z < 0.12$. Whilst \citep{Sureshkumar2024} reports the direct opposite with galaxy mergers preferring low-density environments where high-velocity dispersions are not present. Interestingly, in the middle ground lie \citep{Ostriker1980, McIntosh2008, Perez2009, Darg2010, Alonso2012}, all suggesting that galaxy mergers do not prefer the highest, or lowest densities, but prefer galaxy groups that provide an intermediate density. These regions provide low-velocity dispersions yet with a higher-local density, when compared to isolated field galaxies. In this redshift range, it has also been shown that the relationship between merger fraction and environment may be scale dependant, \citep{Omori2023}. On scales of $0.5$ - $8$ $h^{-1}$ $\mathrm{Mpc}$, merging galaxies are more likely found in low-density environments, however, on reduced scales IllustrisTNG simulation data showed the opposite trend.

As redshift increases, the controversy begins to wane as more studies suggest that galaxy-mergers prefer high-density environments, \citep{Lin2010,Kocevski2011,deRavel2011,Kampczyk2013, Lotz2013,Watson2019,Liu2023,Laishram2024,Shibyua2025}. At $z \approx 1$, \citep{Lin2010,deRavel2011} shows merger rate to be increased in higher density environments, and more recently \citealt{Shibyua2025} showed the highest redshift, $z \approx 2-5$, signal that this relationship continues further back in time. Expectedly, due to the abundance of either type of galaxy in high and low density environments, at $z \approx 1$, \citet{Lin2010} found that gas-rich or wet mergers were more likely to occur in low or intermediate-density regions like the filaments or voids, while gas-poor or dry mergers are more likely in high-density regions. Galaxy mergers, both observationally and in simulations, have been shown to lead to periods of increased star-formation \citep{Ellison2008,Knapen2015,Renaud2022}. Both gas-rich and gas-poor mergers have been shown to lead to increased SFR, \citep{Fensch2017}, however there is conflicting evidence for which of these merger types is the most efficient. \citealt{Fensch2017} finds that lower gas-fraction mergers lead to higher relative increases in SFR when compared to higher-redshift, high gas-fraction mergers which is contrary to the findings of \citealt{Darg2010}. In either case, whilst during the merger metallicity is seen to be reduced, this period of increased SFR leads to an increased rate of chemical enrichment post-merger that may drive galaxies above the MZR \citep{PerezDiaz2024}. In this work, going back to $z = 3.5$ shows gas fractions that are comparable between all three environments, suggesting that gas-rich mergers, which lead to increased SFR and metallicity, could be occurring throughout the Universe. Although this description of mergers can explain an increased metallicity at earlier times for galaxies that have undergone gas-rich mergers, it cannot explain our result that this enrichment occurs for lower mass galaxies, specifically in the node environment, at $z = 3.5$. If mergers were the reason for this early enrichment, then higher-mass galaxies would show increased residuals due to the assembly of mass during the merger. We would also see the signal in all environments due to the presence of gas in all environments at this early time.

Another process that could potentially contribute to this early enrichment is gas accretion, or a lack-thereof. A key difference between cosmic environments is the availability of gas to the galaxies within them. It is shown that satellite galaxies within clusters experience significantly reduced gas accretion rates, leading to a faster quenching of SFR compared to less dense environments \citep{VandeVoort2017}. These galaxies quickly burn through their gas reservoir and with no more pristine gas to accrete, due to several processes that are enhanced within dense environments, they experience environmental starvation. This leads to a halted SFR whilst stellar metallicity continues to increases without further cold-gas accretion \citep{Baker2024}. The rate at which a galaxy can become quenched also depends on its stellar mass, with lower stellar mass galaxies being more susceptible not only to starvation within the cluster \citep{Peng2015}, but to quenching via other cluster-based mechanisms such as ram-pressure stripping (RPS) \citep{GunnGott1972}, which occurs on much shorter timescales than the aforementioned more-drawn out starvation within clusters and likely plays a significant role in the quenching process of galaxies in large groups. \citep{Ditrani2025} recently showed that galaxies in these large groups have shorter star-formation timescales which supports this idea. It has been shown that, whilst RPS will typically lead to a quenched galaxy, SFR is dramatically increased by the presence of RPS in a system, typically in the disk due to the compression of disk gas \citep{Lee2020}, and even in the wake of the stripped galaxy \citep{Kapferer2009}. Boosted SFR in the disk of galaxies undergoing RPS have also been observed in the GASP galaxies, \citep{Vulcani2018}, and multiple individual galaxy studies \citep{Merluzzi2013, Kenney2014}. One can imagine that given sufficient time this enhanced SFR would lead to an increase in $Z_{g}$ due to the initial RPS. 

At $z = 3.5$ we already report a reduction in $\mathrm{log(SFR)}$ of approximately $0.12$ dex, a dramatic reduction in $\Delta f_{\text{gas}}/\Delta t$ and an increased $\Delta Z_{g}/\Delta t$ for satellites surrounding cluster centers, suggesting galaxies in this region are quickly losing their gas, seeing increases in gas metallicity and becoming quenched. This demonstrates a possible galaxy-starvation signal at this early epoch. In groups rather than clusters, \citep{Rhee2024} found that lower-mass, $M_{\star} < 10^{8.2}$ $[M_{\odot}]$, satellite galaxies within the group experienced significant gas loss compared to higher mass galaxies, suggesting that they are more sensitive to quenching mechanisms. As our highest residuals exist for the low stellar mass galaxies in our sample, $M_{\star} < 10^{9.8}$ $[M_{\odot}]$ at $z = 3.5$, following the aforementioned study, one can imagine that this early enrichment could be caused by the starvation of these low mass galaxies in clusters that is occurring on shorter timescales than for the higher stellar mass galaxies that require longer to become quenched. Notably, \citep{Rhee2024} also reported that slightly higher mass  $M_{\star} < 10^{9.1}$ $[M_{\odot}]$ galaxies often experienced a rejuvenation phase later in their infall. As $M_{\star}$ increases with cluster-centric distance \citep{Rowntree2024}, it is possible that the increased SFR in Fig. \ref{fig:dXdt_Radial_Profiles} for low $\mathrm{d_{cluster}}$ could be displaying this rejuvenation in the more massive population of galaxies, post an earlier phase of quenching that leads to the equally interesting dip in SFR in the same figure. 

This notion of cold gas-availability is directly opposite in the void environment, where galaxies typically have more access to the accretion of cold, pristine gas to fuel continued star formation \citep{Bahe2013}. This accretion of pristine gas helps to regulate the gas metallicity of the galaxy \citep{Delucia2020}, keeping it lower than seen for galaxies in more dense environments. The galaxies in these low density environments do not experience this strangulation, and as such do not show residuals from the MZR at $z = 3.5$. We see this signal in Fig. \ref{fig:OFe_LogZ} in which a turnover in $Z_{g}$ exists for low redshift, high mass galaxies in the voids as [O/Fe] decreases. In this same space, the nodes show a dramatically different signal, with no turnover, leading to higher $Z_{g}$ as [O/Fe] decreases. Although this turnover occurs for the lower redshift bins, $z < 2$, and may be unrelated to the early enrichment, it warrants further discussion. The availability of pristine gas in the voids means that the high stellar mass galaxies, with the largest gravitational potentials, can under go accretion events which cause $Z_{g}$ to fall. As this event occurs, star-formation can be triggered, which at first may lead to an increase in [O/Fe] due to the core-collapse of massive stars, however after a delay, the Type 1a SN rate will be increased leading to iron enrichment, decreasing [O/Fe]. In our result, this does not occur in the nodes, suggesting that the lack of this cold-accretion mode removes the regulation of $Z_{g}$, allowing gas metallicity to continue increasing past the maximum observed in other environments. It is clear that this process is most likely to occur for high mass galaxies within low density environments. We would also expect $Z_{g}$ to increase due to these processes however we observe the opposite. We propose that this is due to active cold-accretion in high mass galaxies that keeps reducing $Z_{g}$ and is not reflected in [O/Fe], as [O/Fe] in the gas mainly follows the ratio between the core-collapse and Type Ia SN rates, unless the accretion also involves gas with low [O/Fe], for instance from the ejected ISM of other galaxies, already polluted by Type Ia SNe. Interestingly, the filament galaxies fill an intermediary role, where $Z_{g} / \mathrm{[O/Fe]}$ is steeper in the high mass, high metallicity region than seen for the nodes, however no significant turnover is seen. We can also consider this from the perspective of stellar populations in the Milky Way and the so-called two-infall model \citep{chiappini1997}. This turnover is seen and modelled in \citet{Spitoni2019,spitoni2021} to reproduce the observed dichotomy in the [O/Fe]-[Fe/H] chemical abundance pattern between thick and thin disc stars in the Milky Way (e.g., see \citealt{fuhrmann1998,bensby2003,hayden2015, Snaith2015,Miglio2021,vincenzo2021}) and is related to a second vigorous accretion event of pristine gas that triggered the formation of the thin disc in Milky Way-like spiral galaxies that are more typically found in the field (e.g., see also \citealt{kobayashi2023,Somawanshi2024}). This idea agrees closely with the more abstracted cold-gas accretion explanation between the environments. From this we also anticipate this result is linked to the galaxy morphologies found in these regions; elliptical galaxies are more common in the clusters, which do not show signs of a multi-disc structure, and concurrently do not show a turnover in this space. The void, as mentioned, typically houses more spiral galaxies, and as such this turnover is observed.

A final consideration for this early enrichment in nodes is the presence of AGN in high-density regions. AGN feedback has been suggested to significantly impact SFR and the kinematic properties of galaxies around filaments and nodes, \citep{Kraljic2019}, as such an effected on metallicity could be expected after sufficient time post-SFR increase. AGN-fraction has been shown to decrease with cluster-centric radii for low redshifts, however, the physical number of AGN does increase across the same range \citep{Hwang2012}. Recently, a study of high redshift galaxies, $1 < z < 3$, has shown the AGN-fraction to increase instead for higher-density environments \citep{Gatica2024}. AGN activity helps to push gas out of the host galaxy, moving it towards quiescence and higher metallicities \citep{Peluso2023}. As such, at higher redshifts, a higher AGN-fraction could contribute to a more rapid evolution of gas metallicity in node galaxies, leading to our observed result.
In conjunction with this, within clusters, an anisotropic quenching signal is present that aligns itself with the minor and major axis of the brightest central galaxy. In the past, this has been explained to be due to the presence of the AGN jets along the minor axis, which clear out the intergalactic medium, reducing the ram pressure that galaxies experience in these regions, preserving their SFR \citep{Navarro2021}. Although this process preserves SFR on short time scales, it may lead to the galaxies in the jet region becoming starved of gas after they have consumed what is available to them. Recently, \citealt{Stephenson2024} has shown that this effect extends across multiple Mpc within the cluster, and outside the virial radius of the cluster, with a weak variance in passive galaxy fraction along the major axis. This suggests that the observed effect is actually due to the effect of pre-processing, rather than the AGN feedback. Interestingly, on reduced spatial scales, the presence of these jets have been found to strip cold gas from nearby galaxies, instead helping to quench their SFR \citep{Fujita2008}. Both of these effects show that the presence of AGN activity in galaxy clusters can lead to unique effects on the galaxies within. One could expect that the latter effect is anti-correlated with $d_{cluster}$ due to the local density increase when approaching the cluster core. This idea could lead to, not only the host galaxy, but also the surrounding galaxies showing signs of increased enrichment at early times within clusters. This explanation also agrees with the result that lower mass galaxies are more affected due to their lower gravitational potentials, meaning that these galaxies have their gas stripped more easily by nearby AGN.

To further build on this study, we aim to quantify the effects of galaxy mergers, certain gas-accretion and gas-loss mechanisms and AGNs on the scatter and evolution of the MZR in the 3 distinct global environments within HR5. By providing this analysis, we can determine exactly how these processes directly impact the galaxies in our selection, and how the efficacy of each process varies between nodes, filaments and voids. \cite{2024Singh}, using HR5, studied the effects of environment on AGN activity, as such this work would not only build on this series of papers, but also provide interesting context to other existing work with the simulation. This work would provide a final note on which of these mechanisms are dominating in each environment, and which are most impactful in driving the chemical evolution of galaxies across the universe's lifetime. 
To also add another layer to the further work, we also intend to look into the impact of the commonly used split between passive and star-forming galaxies, along with a split between central and satellite galaxies. Even within the same environment, as seen in \cite{Rowntree2024}, the split between central and satellite galaxies produces dramatic differences in galaxy properties and therefore hints at dramatically different experiences and evolution paths for these galaxies. Due to the commonality of both this split and the significance of the split between passive and star-forming galaxies, understanding how these splits relate to the context of environments, the MZR, and the aforementioned physical processes is key in furthering our understanding of galaxy evolution.
To round out the study, and our results linking to the to the fundamental metallicity relation, we acknowledge the significance of providing a quantitative disentanglement of the contribution to the scatter in the MZR of the galaxy properties that we identify in this study. A viable method of doing this would be to employ machine learning techniques like multi-variate regression to our feature set, ranking each feature by its impact on the scatter in the MZR. This is something that has been carried out in other studies like \cite{Baker2024}, and providing the perspective from HR5 would be a useful contribution to this conversation.
These three ideas for the future study aim to answer the questions left open from the discussion in this study from HR5's perspective.


\section{Conclusion}
\label{sec:Conclusions}

Using the outputs of the Horizon Run 5 simulation from redshift $z=0.5$ to $z=4.5$ we present an analysis of the scatter in the gas MZR and how it evolves in redshift, helping us to better understand where and when enrichment occurs in the Universe. By quantifying three global environments, nodes, filaments and voids, based on the structure estimates created by the {\tt T-ReX} filament finder and the galaxy cluster catalogues from HR5 at each snapshot, we define the environment that each galaxy exists within, studying the environmental contribution to the scatter in the MZR at different points during the evolution of the Universe. We also investigate the galaxies that exist in the physical space surrounding the cores of nodes and filaments using the perpendicular distance to filament, $d_{\text{skel}}$, and the cluster-centric distance, $d_{\text{cluster}}$, studying the radial profiles of gas metallicity, gas fraction and [O/Fe] to further build on how these properties are spatially distributed. Finally, we calculated the rate of change of gas metallicity, $\Delta Z/ \Delta t$, and gas fraction, $\Delta f_{\text{gas}}/\Delta t$, providing insight into the instantaneous evolution of galaxies in the three environments. From this analysis, we concluded the following;

\begin{enumerate}

\item Deviations from the total gas MZR due to the environment emerge at different points during the evolution of the Universe for different environments, with the nodes showing accelerated chemical enrichment. Low-$M_{\star}$ galaxies in the nodes, as early as $z = 3.5$, show increased residuals when the other environments show no signs of enrichment. By $z = 2$, low-$M_{\star}$ galaxies in the filaments and voids show slight enrichment whilst the node populations enrichment continued to evolve, extending to higher stellar masses and even larger residuals for the low-$M_{\star}$ population. By $z = 0.5$ all environments show significant deviation from the total MZR, across wider ranges of $M_{\star}$ than for higher redshift snapshots.

\item Low $M_{\star}$ galaxies deviate from the MZR at earlier times than high $M_{\star}$ galaxies. This trend occurs in all environments where the lowest-$M_{\star}$ galaxies demonstrate chemical enrichment first, with each consecutive snapshot showing a more massive population becoming enriched as you move along the MZR. Interestingly, these low $M_{\star}$ galaxies in the nodes seem to reach a residual saturation value by $z = 1$, after which they see no further enrichment whilst the higher $M_{\star}$ bins continue to move towards this value. This process as a whole demonstrates a "peeling" of each environment away from the MZR with the passage of time, with the low $M_{\star}$ galaxies starting this action.

\item $\mathrm{Z_{g}/[O/Fe]}$ evolves with redshift differently in nodes, filaments and voids. The trend from $z = 4.5$ to $2.5$ shows the same steepening in all three environments with a lower [O/Fe] being associated with higher $Z_{g}$ and $M_{\star}$, however, after $z = 2.5$ the environments deviate from one another. $\mathrm{[O/Fe]/Z_{g}}$ in the nodes flattens for low redshifts, $z < 1$, with lower values of [O/Fe] leading to continuously higher values of $Z_{g}$. The filaments and voids do not flatten after $z = 2.5$ and instead continue to steepen, showing significantly lower $Z_{g}$ for the same [O/Fe] values as the nodes. Finally, the voids mimic the filaments steepening, however show a dramatic turnover at higher $Z_{g}$, after which a lower [O/Fe] leads to a lower $Z_{g}$ and yet still a higher $M_{\star}$.
This suggests that void galaxies with high $M_{\star}$ in the simulation underwent larger accretion rates of pristine gas or gas-rich mergers that diminished their average $Z_{g}$. This process can leave an imprint on the [O/Fe]/[Fe/H] abundance patterns of the stellar populations of massive void galaxies; this may be explored in future work.

\item Galaxies in all environments that are either rapidly gaining or losing gas, for all snapshots, demonstrate increased gas metallicity, $Z_{g}$, relative to the total MZR. Between the two cases that demonstrate increased residual, the highest residuals occur for galaxies that are rapidly losing gas at $z = 0.5$. High residual galaxies, from both cases, are associated with an instantaneously increasing gas metallicity, $\Delta Z_{g}/\Delta t$, low SFR, and typically are found in nodes. In conjunction, galaxies that are stable in $\Delta f_{\text{gas}}/\Delta t$ show the lowest residuals for all snapshots. At $z = 0.5$, this low residual, more gas-stable population is coupled with very low $\Delta Z_{g}/\Delta t$, higher SFRs and is most likely found in the void or possibly filaments.

\item Galaxies within the centers of nodes/clusters are losing gas significantly faster than galaxies in the outskirts of these structures or in and surrounding filaments. Concurrently, 
independent of redshift, at a given  $d_{\text{cluster}}$, all galaxies will experience a similar value of $\Delta f_{\text{gas}}/\Delta t$. This value of $\Delta f_{\text{gas}}/\Delta t$ drops dramatically with smaller $d_{\text{cluster}}$.

\item When binned by $\log(\text{SFR})$, the slope of the MZR changes depending on the environment in question. Nodes demonstrate a steepening of the MZR, where higher $M_{\star}$ galaxies show higher residuals from the total MZR, the opposite that is observed in Fig. \ref{fig:MZR_Scatter_Redshifts}. The filament environment shows a slope most similar to that of the total MZR, whilst the void galaxies remain to show lower $M_{\star}$ galaxies to have the highest magnitude of residual. The slope of $\log(\text{SFR})$ against the residual, $d\log(Z_{g}/Z_{\sun})$, is also dependent on environment and redshift. Across the full range of SFR in the nodes, we see a large variance in residual at $z = 0.5$. In the filaments and voids, at the same redshift, this variance has halved. For higher redshifts, this slope is steepened in all three environments. 

\end{enumerate}

Our conclusions show when and where the enrichment of galaxies occurs in the HR5 simulation's volume based on our cosmic structure definitions, providing great insight into where we could expect to see the early enrichment of galaxies occurring in our own Universe. They also provide a perspective into the possible different processes occurring in different cosmic environments that may lead to this accelerated enrichment in dense regions that we report, suggesting possible explanations for the observation of high metallicity galaxies at earlier times. These results combine in such a way that paints a story that galaxies and their evolutions must not only be considered as individual bodies but as parts of a fully interconnected system, as the interaction between a galaxy, its neighbours and the surrounding environment is key in the acceleration of their evolution, and a deciding factor in how they come to rest at late times. With the advent of equally deep and wide surveys like DESI \citep{DESI2023}, Euclid \citep{Euclid2022} and PRIMA \citep{Glenn2025}, spectroscopic measurements of galaxies along with their surrounding environment will become publicly available, allowing these findings to provide inspiration for how these surveys can be used to further our understanding of early galaxy enrichment, global environments, and galaxy evolution as a whole.
In the future, we hope to further explore the relationship between [O/Fe]/[Fe/H] and environment with the hopes of deepening our understanding of how gas-accretion and galaxy mergers impact the chemical evolution of galaxies. It is also of interest to explore the morphology of galaxies in these regions, relating them to their gas-accretion history and the two-infall model. Signatures of cold-accretion may be left in the abundance patterns of galaxies, which may relate to the importance of environment on forming certain morphologies of galaxy.

\section*{Acknowledgements}

We acknowledge the support of STFC (through the University of Hull's Consolidated Grant ST/R000840/1) and ongoing access to {\tt viper}, the University of Hull High-Performance Computing Facility. FV thanks the National Science Foundation (NSF, USA) under grant No. PHY-1430152 (JINA Center for the Evolution of the Elements). We acknowledge the support of computing resources at the Center for Advanced Computation (CAC) at KIAS. AS was supported by KIAS Individual Grants (PG080901) and acknowledges support from the UK Research and Innovation (UKRI) Nottingham Astronomy consolidated grant ST/X000982/1, Astronomy and Astrophysics at the University of Nottingham - 2023 to 2026 (PI: Simon Dye). J.L. is supported by the National Research Foundation (NRF) of Korea grant funded by the Korea government (MSIT, RS-2021-NR061998). J.L. also acknowledges the support of the NRF of Korea grant funded by the Korea government(MSIT, 2022M3K3A1093827). Large data transfer was supported by KREONET, which is managed and operated by KISTI. This work is also supported by the Center for Advanced Computation at Korea Institute for Advanced Study, (KSC-2018-CHA-0003, KSC-2019-CHA-0002). This work benefited from the outstanding support provided by the KISTI National Supercomputing Center and its Nurion Supercomputer through the Grand Challenge Program (KSC-2018-CHA-0003, KSC-2019-CHA-0002). ONS acknowledges support from ERC Consolidator Grant "GAIA-BIFROST" (Grant Agreement ID 101003096) and the UK Science and Technology Facilities Council (Consolidated Grant ST/V000721/1; Small Award ST/Y002695/1. Y.K. is supported by KISTI under the institutional R\&D project (K25L2M2C3). CP is partially supported by the grant SEGAL ANR-19-CE31-0017 of the French Agence Nationale de la Recherche. We thank the reviewer for their comments and insights that significantly improved the work.

\section*{Data Availability Statement}

The data underlying this work is able to be shared upon reasonable request.

\bibliographystyle{mnras}
\bibliography{references}

\appendix

\section{Evolution of Filament Radius in Redshift}

As mentioned in the work, the large-scale structure in the Universe evolves with redshift, typically increasing in size as time passes. This means that the median thickness, or radius, of filaments will increase with lower redshifts. To account for the evolution of structure in this study we mainly focus on our filament finder, {\tt T-ReX}, and the specific galaxy distribution we provide it at each epoch to ensure our computed skeletons, see Fig. \ref{fig:Skeleton_Comparison}, are comparable across time. Although this process is done to account for the evolution of structure, a possible caveat emerges from using a static filament radius, of 1 cMpc, for our filament galaxy definition. To ensure that our results are not significantly effected by this variable we re-produce Fig. \ref{fig:MZR_Scatter_Redshifts}, implementing a quick evolution of filament radius at each epoch which is outlined in table \ref{table3}.

\begin{table}
    \centering
    \caption{The evolution of the filament radius definition that selects galaxies nearby this environment at different redshifts.}
    \begin{tabular}{ |c|c| }
     z & Filament galaxy definition [cMpc]  \\ 
     \hline
     0.5 & 1.00 \\  
     \hline
     1.0 & 0.85 \\
     \hline
     1.5 & 0.70 \\
     \hline
     2.0 & 0.55 \\
     \hline
     3.5 & 0.4 \\
     \hline
     4.5 & 0.25 \\
     \hline
    \end{tabular}
    
    \label{table3}
    \end{table}

\begin{figure*}
    \centering
    \includegraphics[scale=0.65]{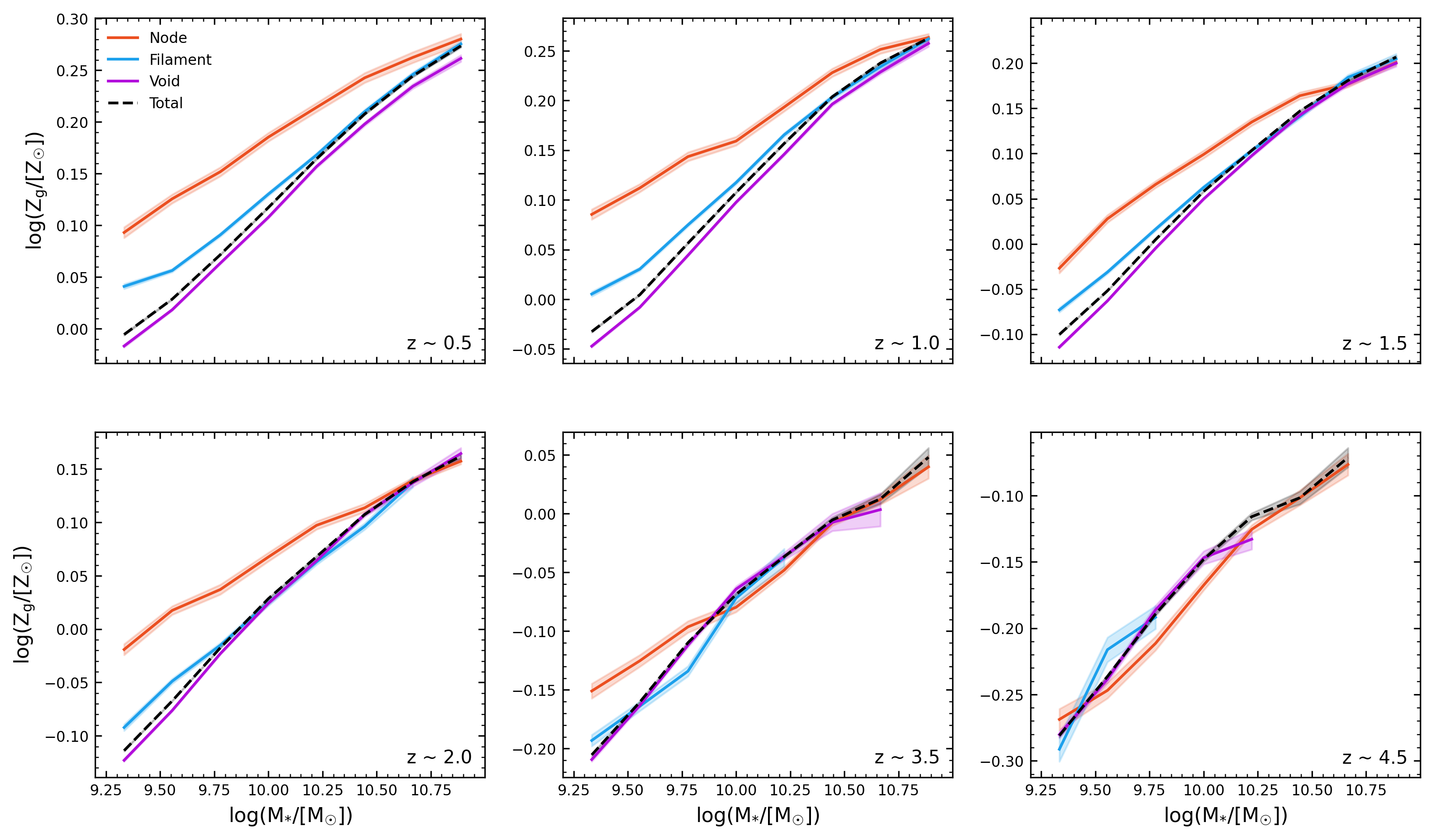}
    \caption{The MZR for node galaxies (orange), filament galaxies (blue), void galaxies (purple) and the total population (black dashed), over 10 consecutive, equal-sized bins in $M_{\star}$, for 6 of the chosen redshift snapshots, where bins containing less than 15 galaxies have been removed. In this figure, the filament population is populated using the variable radius for each snapshot outlined in table \ref{table3}. The shaded regions represent the standard error on the median.}
    \label{fig:MZR_Redshifts_Filament_Radius_Evolution}
\end{figure*}

By comparing Fig. \ref{fig:MZR_Redshifts_Filament_Radius_Evolution}, to Fig. \ref{fig:MZR_Scatter_Redshifts}, we do not observe a major change in the results. The only observed change is that the median filament MZR in all bins of $M_{\star}$, and at all epochs, is marginally closer to the MZR for the total population of galaxies, but not by an amount that affects the studies outcome. Although the latter is true, in future studies we will opt to include this evolution of filament radius as it better reflects the physical nature of these structures as they evolve in time.

\end{document}